\documentclass[reprint, amsmath,amssymb,showpacs,floatfix, aps, prb, twocolumn, superscriptaddress,longbibliography]{revtex4-2}
\usepackage{graphicx}
\usepackage{dcolumn}
\usepackage{bm}
\usepackage{color}
\usepackage{hyperref}
\usepackage[normalem]{ulem}
\hypersetup{colorlinks=true,citecolor=blue,linkcolor=blue, urlcolor=blue}
\newcolumntype{M}[1]{>{\centering\arraybackslash}m{#1}}
\newcolumntype{N}{@{}m{0pt}@{}}
\def\nn{\nonumber}
\def\({\left(}
\def\){\right)}
\def\[{\left[}
\def\]{\right]}
\def\a{\alpha}
\def\b{\beta}
\def\d{\delta}

\def\e{\epsilon}

\def\s{\sigma}

\def\w{\omega}

\def\g{\gamma}

\def\bfB{\mathbf{B}}

\def\bfb{\mathbf{b}}

\def\bfk{\mathbf{k}}
\def\bfm{\mathbf{m}}

\def\bfq{\mathbf{q}}

\def\bfR{\mathbf{R}}

\def\bfx{\mathbf{x}}

\def\bfS{\mathbf{S}}

\def\>{\rangle}
\def\<{\langle}

\begin{document}

\title{First-Principles Approach to Spin Excitations in Noncollinear Magnetic Systems}

\author{Hsiao-Yi Chen}
\affiliation{Institute for Materials Research (IMR), Tohoku University, Sendai, 980-8577, Japan}
\affiliation {RIKEN Center for Emergent Matter Science (CEMS), Wako 351-0198, Japan}

\author{Ryotaro Arita}
\affiliation {RIKEN Center for Emergent Matter Science (CEMS), Wako 351-0198, Japan}
\affiliation {Department of Physics, The University of Tokyo, Bunkyo-ku, Tokyo 113-0033, Japan}

\author{Yusuke Nomura}
\affiliation{Institute for Materials Research (IMR), Tohoku University, Sendai, 980-8577, Japan}
\affiliation{Advanced Institute for Materials Research (WPI-AIMR), Tohoku University, Sendai, 980-8577 Japan}

\date{\today}

\begin{abstract}
We present a first-principles method based on density functional theory and many-body perturbation theory for computing spin excitations in magnetic systems with noncollinear spin textures. Traditionally, the study of magnetic excitations has relied on spin models that assume magnetic moments to be localized. Beyond this restriction, recent $ab~initio$ methods based on Green’s functions within the local spin-density approximation have emerged as a general framework for calculating magnetic susceptibilities. However, their application has so far been largely limited to collinear ferromagnetic and antiferromagnetic systems. 
In this work, we extend this framework and enable the treatment of large-scale noncollinear magnetic systems by leveraging a Wannier-basis representation and implementing an ansatz potential method to reduce computational cost. 
We apply our method to the spin-spiral state of LiCu$_2$O$_2$, successfully capturing its steady-state spin-rotation pitch in agreement with the experimental measurement and resolving the characteristic magnon dispersion. We further analyze the interplay between the spiral spin structure and the on-site spin-exchange splitting, and elucidate the crucial role of magnetic dipoles on ligand ions in mediating effective ferromagnetic interaction among the primary spins on $\rm Cu^{2+}$ ions. Finally, we provide a theoretical prediction of the magnon dispersion on top of the helical spin background in high agreement with the experimental measurement. Overall, this work establishes a general and computationally efficient framework for simulating collective spin dynamics in noncollinear magnetic systems from first principles, exemplified by—but not limited to—spin-spiral states.
\end{abstract}

\maketitle

\section{Introduction \label{Sec:Intro}}
\vspace{-10pt}
Noncollinear magnetic materials, characterized by spins aligned along different axes in the ground state, have been recognized as promising platforms for next-generation spintronic applications \cite{brataas2006non,nan2020controlling,hirohata2020review,rimmler2025non}. Their intrinsic noncollinearity gives rise to complex spin textures such as skyrmions and spin helices, which are topologically distinct from collinear configurations and are considered strong candidates for robust, high-density magnetic memory devices \cite{zhang2020skyrmion,tokura2010multiferroics}.
Furthermore, they combine advantages of both ferromagnets and antiferromagnets: they exhibit faster magnetic dynamics and are free from stray field effects \cite{jungwirth2016antiferromagnetic,baltz2018antiferromagnetic}, while simultaneously supporting spin-polarized currents and hosting exotic transport phenomena, including the anomalous Hall and Nernst effects \cite{nakatsuji2015large,ikhlas2017large}. 
In addition, noncollinear spin structures can induce the topological Hall effect even in the absence of spin–orbit coupling, which is typically required for the conventional anomalous Hall effect \cite{takagi2023spontaneous,watanabe2024symmetry}.
These features make noncollinear magnets fertile ground for exploring and controlling spin degrees of freedom.
\\
\indent
In parallel, alternative technologies for manipulating and transferring information encoded in spin have evolved significantly over the past decade. Methods using the spin waves---and their quantum counterparts, magnons---to  process data have led to the development of fields of {\it magnonics} \cite{kruglyak2010magnonics,lenk2011building,barman20212021} and of {\it magnon spintronics}, which explore their potential in spin-based devices \cite{stamps20142014,chumak2015magnon}. Compared to the traditional methods that rely on electric currents, magnons enable information transfer without physical particle motion, minimizing energy loss due to Joule heating. In addition, their high speed dynamics offer a pathway to ultrafast spin control \cite{kampfrath2011coherent,li2020spin}, while the corresponding short wavelengths provide the possibility to miniaturize devices down to atomic scales \cite{heinz2020propagation}. More recently, magnons have also garnered interest for their potential applications in quantum information technologies  \cite{yuan2022quantum}.
\\
\indent
Compared to the emerging research on magnon-based applications, understanding the physical properties of spin waves has been a long-standing topic of study since the proposal by Bloch \cite{bloch1930theorie}. The use of localized spin models based on the Heisenberg Hamiltonian \cite{holstein1940field} have proven to be a reliable framework for investigating magnon behavior and remains the predominant approach for describing spin systems and their excitations. While the original development is restricted to collinear magnets, recent efforts have extended spin-wave theories to noncollinear systems in response to the increasing discovery of noncollinear magnetic structures \cite{akagi2013effect,dos2018spin}. 
\\
\indent
To enable direct comparison with experimental measurements and to predict magnetic excitations in new materials, an $ab~initio$ method for simulating spin waves is indispensable. Several theoretical approaches have been proposed to bridge the density functional theory (DFT) with the magnetic force theorem \cite{durhuus2023plane}, or employing local force methods \cite{oguchi1983magnetism,szunyogh2011atomistic,nomoto2020local,hatanaka2024calculation} for estimating the exchange couplings in the spin models from first principles. While these spin model-based approaches are computationally efficient, they 
%are oversimplified and 
lack several essential physical features. 
In particular, they rely on the assumption of localized magnetic moments and thus fail to accurately describe itinerant magnets, which results in large errors in $d$-electron system. Moreover, complex magnetic compounds consisting of multiple spins with varying magnitudes require more sophisticated treatment than the simplified half-integer spin model. These limitations underscore the need for a fully first-principles approach.
\\
\indent
Currently, there are two major approaches for simulating excited states in real material systems from first principles within the framework of DFT beyond the spin model. Time-dependent DFT (TDDFT) can adiabatically account for the temporal evolution of spin excitations under an external field and facilitates the calculation of the magnetic susceptibilities in ferromagnets and antiferromagnets \cite{skovhus2021dynamic,skovhus2022magnons,gorni2022turbomagnon}. However, TDDFT is limited by its reliance on the local density approximation (LDA) for the exchange-correlation kernel while the nonlocal effect is absent. On the contrary, the combination of the DFT with the many-body perturbation theory (MBPT) \cite{aryasetiawan1999green,karlsson2000many,karlsson2000spin} using the Green's function method to incorporate the spin-wave kernel within the GW approximation can correctly include the nonlocal effect by solving the Bethe-Salpeter equation (BSE). Recent developments have extended and refined the discussion of magnon under this framework. Şaşıoğlu $et~al.$ employed a localized Wannier basis to encode the scattering kernel, significantly reducing computational cost and enabling applications to more complex systems \cite{sasioglu2010wannier}. Olsen provided a unified treatment of magnons and excitons by converting the BSE into an effective Hamiltonian \cite{olsen2021unified}, allowing direct access to magnon wave functions. Le $et~al.$ further utilized this wave function to investigate magnon–phonon interactions \cite{le2025magnon}.  Nevertheless, despite these advances, first-principles approaches have so far been limited to collinear magnetic systems and a corresponding framework applicable to noncollinear spin systems has remained absent, leaving a critical gap in the theoretical toolkit.
\\
\indent
In this work, we aim to fill this gap by developing a first-principles method to predict spin-wave excitations in noncollinear magnets. Our approach builds on the MBPT framework 
incorporating several techniques to overcome computational difficulties inherent in complex noncollinear spin systems. 
First, we employ a localized Wannier basis~\cite{aryasetiawan1999green,sasioglu2010wannier} to significantly improve the efficiency of computing the BSE interaction kernel. In addition, by introducing local quantization axes, we can straightforwardly account for spin-flip processes associated with  general spin orientations.
Another challenge arises from the DFT calculations as noncollinear spin textures that span multiple unit cells, such as spin spirals and skyrmion crystals, render conventional DFT calculations prohibitively expensive. To circumvent this computational bottleneck, we employ  the ansatz potential method \cite{chen2025topological}, which can construct the magnetic component of the exchange potential without the high-cost self-consistent procedures.
%implemented within a localized Wannier basis \cite{aryasetiawan1999green,sasioglu2010wannier}, which we extend accommodate systems with general spin orientations. For noncollinear spin textures that span multiple unit cells—such as spin spirals and skyrmion crystals—which are not directly accessible via conventional DFT, we employ the ansatz potential method \cite{chen2025topological} to construct the magnetic component of the exchange potential. This enables the generation of a tight-binding Hamiltonian that faithfully captures the spin structure of the target system and serves as a basis for computing its spin-wave excitations.
\\
\indent
Utilizing the implementation, we apply the formalism to study the helimagnetic state of the quasi-one-dimensional type-II multiferroic, $\rm LiCu_2O_2$. Our results reproduce the experimentally observed spin-helix structure and reveal the underlying microscopic interplay among magnetic atoms.  We further provide first-principles predictions of the spin-wave spectrum atop the helical ground state in high agreement with the experimental measurement. Overall, this work introduces a computational framework broadly applicable to noncollinear magnetic systems and offers a practical tool for analyzing spin excitations and exchange interactions from a unified, first-principles perspective.
\\
\indent
The structure of this paper is as follows. In Sect.~\ref{Sec:Theory}, we present the theoretical formalism employed in this study, including the derivation of the magnetic susceptibility and spectral function for general spin systems, along with their implementation using Wannier functions and the ansatz potential technique. 
In Sect.~\ref{sect:application}, we apply the developed formalism to  $\rm LiCu_2O_2$ and present the corresponding numerical results, together with the analysis of the underlying physical origin. 
In Sect.~\ref{sect:conclusions}, we summarize our findings and outline directions for future research. Technical details of the numerical implementation are provided in the Appendices.

\section{Theoretical methods \label{Sec:Theory}}
\vspace{-10pt}
In this section, we introduce a first-principles framework for computing magnon spectra in magnetic systems with noncollinear spin textures. 
We begin in Sect.~\ref{subsect:MBPT-mag} by deriving the formalism for calculating the magnetic response of a general magnetic system within the MBPT framework. In Sect.~\ref{subsect:wannier-basis}, we reformulate the expressions using a Wannier orbital basis to enable practical computation. Last, in Sect.~\ref{Subsec:AP-method}, we review the ansatz potential method and describe how it can be incorporated into the spin excitation formalism to enable applications to large-scale noncollinear spin structures.

\subsection{Many-body perturbation theory approach for spin excitation in noncollinear magnetic system \label{subsect:MBPT-mag}}
\begin{figure}[t]
    \centering
    \includegraphics[width=0.9\linewidth]{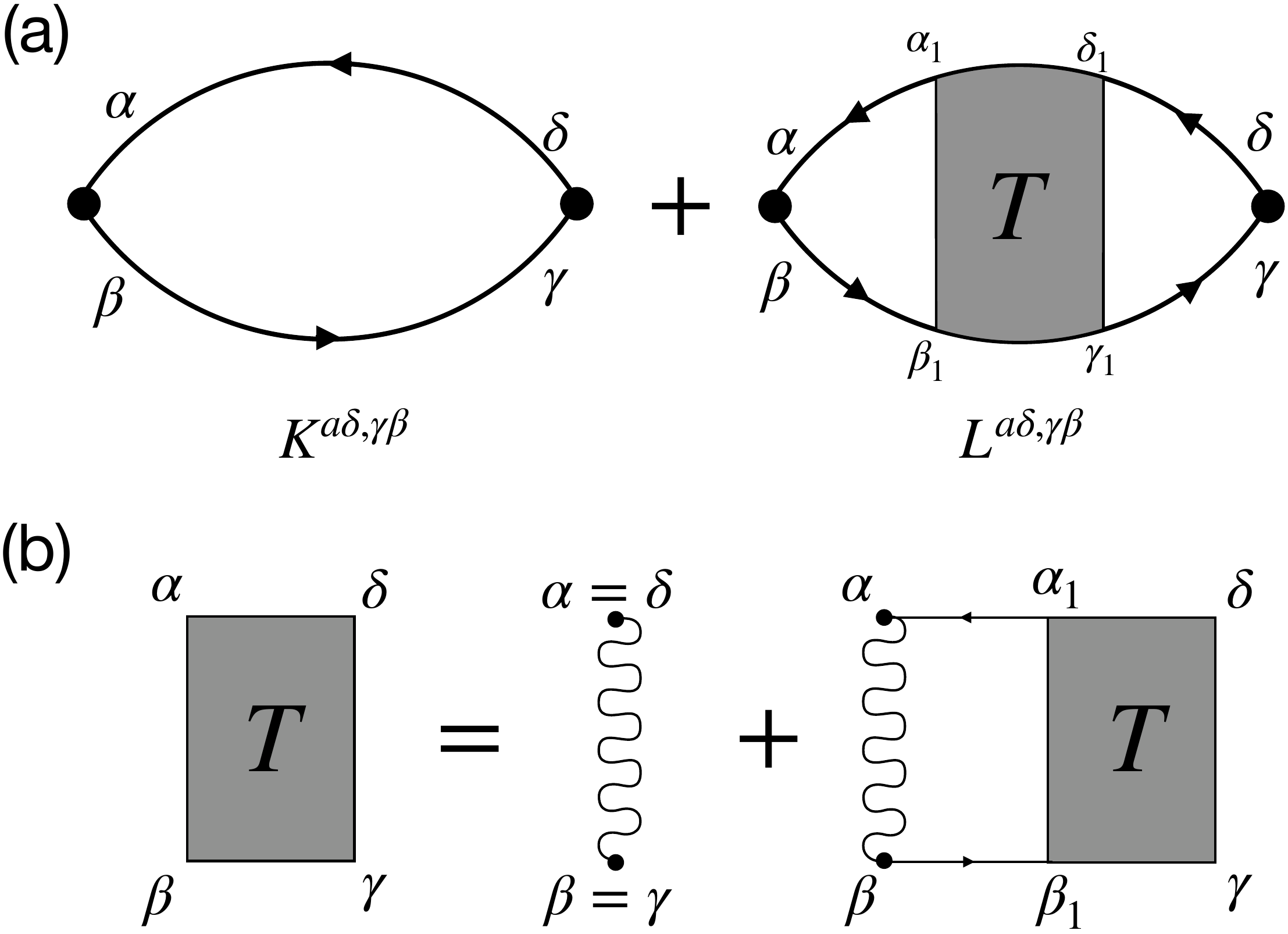}
    \caption{Diagrammatic representation of (a) magnetic response function and (b) scattering kernel in the form of BSE. Compared to the collinear case in Ref.~\cite{sasioglu2010wannier}, we include more spin index to take account for the intrinsic spin rotation in the non-interacting Green's function, for which $G^{0}_{\a\b}$ holds finite off-diagonal component under noncollinear magnetic potential.}
    \label{Fig:diagram}
\end{figure}

The MBPT formalism for simulating spin excitations in general spin systems starts with a Hamiltonian:
\begin{eqnarray}
    H=H_0+H_{\rm int}+ {\boldsymbol{ \sigma}}\cdot \bfB
\end{eqnarray}
where $H_0$ is the solvable non-interacting part, $H_{\rm int}$ is the interaction Hamiltonian, and $\bfB$ is the external magnetic perturbation coupled to electrons via the spin operator ${\boldsymbol{ \sigma}}$, the dynamical spin susceptibility is defined as the linear response of the spin operator to the external perturbation:
\begin{eqnarray}
    R^{ij}(1,2)=\frac{\d\<\hat{\s}^i(1)\>}{\d B_j(2)}=-i\<T\left[\hat{\sigma}^i(1),\hat{\sigma}^j(2)\right]\>
    \label{Eq:Rij}
\end{eqnarray}
where the indices $i,~j=x,y,z$ refer to Cartesian components in the laboratory frame, $T$ denotes the time-ordering operator, and the shorthand notation $1=(t_1,\bfx_1)$, $2=(t_2,\bfx_2)$ is used for space-time coordinates. The spin expectation value is computed by 
\begin{eqnarray}
    \<\hat{\s}^i(1)\>=\sum_{\a\b}{\sigma}_{\a\b}^i\<T\left[\hat{\psi}_\a(1)\hat{\psi}^\dagger_\b(1^+)\right]\>=\sum_{\a\b}{\sigma}_{\a\b}^iG_{\a\b}(1,1^+)\nn\\
    \label{Eq:sigma_1}
\end{eqnarray}
where $\hat{\psi}^{(\dagger)}$ is the electron annihilation (creation)  operator with Greek symbol denoting the spin component and $G_{\a\b}$ is the spin-dependent Green's function.
\\
\indent
Due to the existence of the $H_{\rm int}$, the Green's function acquires the self-energy correction which can be expressed in terms of the Dyson equation:
\begin{eqnarray}
    &&G_{\a\b}(1,2)=G^0_{\a\b}(1,2)\nn\\&&~~~~~~+\sum_{\g\d}\int_{34}G^0_{\a\g}(1,3)\Sigma_{\g\d}(3,4)G_{\d\b}(4,2)
    \label{Eq:Dyson}
\end{eqnarray}
where $G^0_{\a\b}$ is the non-interacting  Green's function associated with $H_0$. We follow Ref.~\cite{sasioglu2010wannier} and focus on the electron-electron Coulomb interaction as the dominant contribution to $H_{\rm int}$, for which the GW-approximation is employed to express the self-energy as:
\begin{eqnarray}
    \Sigma_{\a\b}(1,2)=iG_{\a\b}(1,2)W(1,2)
    \label{Eq:GW}
\end{eqnarray}
where $W$ is the screened Coulomb interaction.
\\
\indent
Following the algebraic detail of the derivation for the collinear case in Ref.~\cite{sasioglu2010wannier}, the expression of the magnetic susceptibility in a general noncollinear system can be obtained by combing Eqs.~(\ref{Eq:sigma_1}-\ref{Eq:GW}), yielding \footnote{The results reduce to the expression in Ref.~\cite{sasioglu2010wannier} in the collinear limit by taking $G^0_{\a\b}=G^0\d_{\a\b}$.} :
\begin{eqnarray}
    &&R^{ij}(1,2)=-\sum_{\a\b\g\d}\s^i_{\b\a}\s^j_{\g\d}\nn\\
    &&~~~~\left[K^{\a\d,\g\b}(1,2;2,1)+L^{\a\d,\g\b}(1,2;2,1)\right].
    \label{Eq:Rij}
\end{eqnarray}
The first term:
\begin{eqnarray}
    &&K^{\a\d,\g\b}(1,3;4,2)=i G_{\a\d}(1,3)G_{\g\b}(4,2)\nn\\
    &&~~~~~~~~~~=
    \frac{1}{N_\bfk N_{\bfk'}}
    \sum_{mn,\bfk\bfk'}
    \frac{
    (f_{m\bfk'}-f_{n\bfk})
    }{\w-(\e_{n\bfk}-\e_{m\bfk'})+i\eta}\nn\\
    &&~~~~~~~~~~~~~~~\times 
    \psi^\a_{n\bfk}(1)
    \psi^{\d*}_{n\bfk}(3)
    \psi^{\g}_{m\bfk'}(4)
    \psi^{\b*}_{m\bfk'}(2)
    \label{Eq:Kab}
\end{eqnarray}
corresponds to the non-interacting two-particle Green's function. This term captures the contribution from the Stoner excitations, which represent incoherent spin-flip transitions occurring in the absence of many-body interactions.
\\
\indent
On the other hand, the second term in Eq.~(\ref{Eq:Rij}) takes the form:
\begin{eqnarray}
    &&L^{\a\d,\g\b}(1,2;2,1)=
    \int_{3456} K^{\a\a_1,\b_1\b}(1,3;4,1)\nn\\
    &&\times T^{\a_1\d_1,\g_1\b_1}(3,5;6,4) K^{\d_1\d,\g\g_1}(5,2;2,6)
    \label{Eq:Lab}
\end{eqnarray}
where $T^{\a_1\d_1,\g_1\b_1}$ is the interaction kernel that accounts for the scattering interaction expressed by the Bethe-Salpeter equation (BSE):
\begin{eqnarray}
    &&T^{\a\d,\g\b}(1,3;4,2)=W(1,2)\d_{\a\d}\d_{\g\b}\d(1-3)\d(2-4)
    \nn\\
    &&+W(1,2) \int_{56} K^{\a\a_1,\b_1\b}(1,5;6,2)T^{\a_1\d,\g\b_1}(5,3;4,6)\nn\\
    \label{Eq:BSE}
    \end{eqnarray}
which sums over the ladder diagrams of the screened Coulomb interaction, $W$. While the Coulomb interaction contributes to spin-non-flipping channel, it correlates the electron wave function of different space, making up the collective effect of excitation which corresponds to the spin wave (magnon). Overall, the results can be summarized in diagrams as shown in Fig.~\ref{Fig:diagram}. It is worthy to note that the noncollinear expression presented here is consistent with the collinear formalism in Ref.~\cite{aryasetiawan1999green,sasioglu2010wannier}, where the spin-up channels and spin-down channels are isolated such that the off-diagonal component in the Green's function vanishes. Mathematically, the collinear result can be recovered by contracting the spin index:
\begin{eqnarray}
    &&G^0_{\a\b}\rightarrow G^0_{\a\a}=G^0_{\a}\nn\\
    &&K^{\a\d,\g\b}\rightarrow K^{\a\a,\b\b}=K^{\a\b}\nn\\
    &&T^{\a\d,\g\b}\rightarrow T^{\a\a,\b\b}=T^{\a\b}\nn\\
    &&L^{\a\d,\g\b}\rightarrow L^{\a\a,\b\b}=L^{\a\b}
\end{eqnarray}
while in Fig.~\ref{Fig:diagram}, the two ends of the electron propagator are reduced to share the same spin index.
\\
\indent
The expressions derived in Eqs.~(\ref{Eq:Rij}–\ref{Eq:BSE}) 
combined with DFT calculations
establish a general framework for computing the magnetic susceptibility in noncollinear systems. 
However, the computation of spin excitations in noncollinear spin systems can be overwhelmingly demanding both in MBPT and DFT parts.
To overcome the difficulties, we introduce numerical techniques: 
\begin{itemize}
    \item {\it Solving the BSE in Wannier basis}: the four-point space–time representation of BSE imposes significant computational demands, making direct numerical implementation impractical for large or complex systems.  To make the formalism tractable and applicable to complex systems, in the following, we reformulate the result using a Wannier basis to reduce the computational cost associated with the space–time indices \cite{sasioglu2010wannier}.
    Furthermore, Wannier basis enables the assignment of local quantization axes at each magnetic site, which makes it straightforward to implement spin-flip excitations in BSE. 
    \item {\it Ansatz potential method for large-scale DFT}: Many noncollinear spin textures extend over multiple unit cells, rendering standard DFT calculations computationally prohibitive. To address this limitation, we incorporate the ansatz potential method into the framework \cite{chen2025topological}, which enables the construction of large-scale noncollinear exchange potentials without the need for self-consistent field calculations.  
\end{itemize}
We discuss these techniques in detail in the following subsections.
%However, their formulation in the four-point space–time representation imposes significant computational demands, making direct numerical implementation impractical for large or complex systems.  To make the formalism tractable and applicable to complex systems, in the following, we reformulate the result using a Wannier basis to reduce the computational cost associated with the space–time indices  \cite{sasioglu2010wannier}. In addition,  many noncollinear spin textures extend over multiple unit cells, rendering standard DFT calculations computationally prohibitive. To address this, we further incorporate the ansatz potential method into the framework \cite{chen2025topological}, which enables the construction of large-scale noncollinear exchange potentials without requiring self-consistent field calculations.  

\subsection{Efficient MBPT implementation using Wannier basis \label{subsect:wannier-basis}}

\begin{figure}[t]
    \centering
    \includegraphics[width=0.9\linewidth]{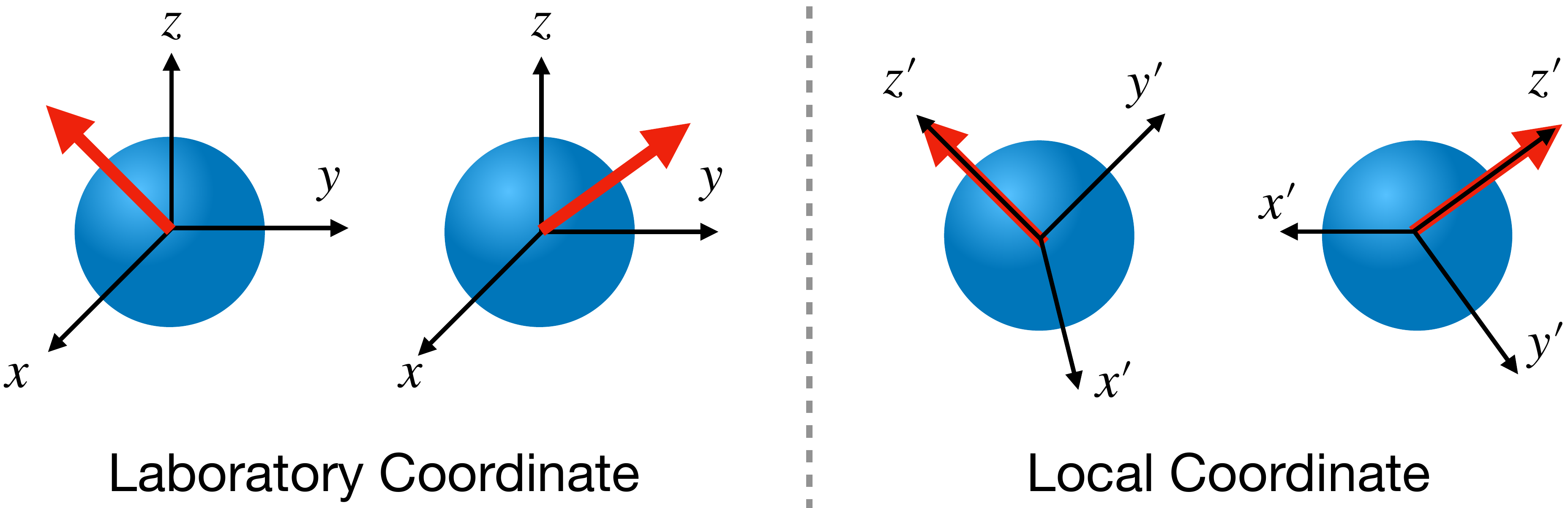}
    \caption{Instead of laboratory coordinate (left), we choose the local coordinate by setting the $z'$-axis to point along the local spin moment such that the spin-flipping process is well-defined in the local frame (right).}
    \label{Fig:local}
\end{figure}

The computational challenges in evaluating Eqs.~(\ref{Eq:Lab}) and (\ref{Eq:BSE}) poses significant challenges to numerical implementation.  This complexity can be substantially reduced by adopting the localized Wannier basis \cite{sasioglu2010wannier}. In this work, to accommodate systems with arbitrary spin orientations, we adopt a spin-symmetric Wannier basis,  in which each Bloch wave function can be expanded as:
\begin{eqnarray}
    &&\psi_{m\bfk}(\bfx)=
    \begin{pmatrix}
        \psi^{u}_{m\bfk}\\\psi^{d}_{m\bfk}
    \end{pmatrix}
    \nn\\
    &&
    =\sum_{\bfR,N}e^{i\bfk\cdot\bfR}
    w_{N\bfR}(\bfx)
    \left[
    U^{\uparrow_L}_{Nm\bfk}
    \chi^{\uparrow_L}_N
    +
    U^{\downarrow_L}_{Nm\bfk}
    \chi^{\downarrow_L}_N
    \right]
    \label{Eq:wan_expand}
\end{eqnarray}
where $\chi^{\uparrow_L}_N$ and $\chi^{\downarrow_L}_N$ are the two-component spins whose quantization axis is aligned with the local spin moment ($\uparrow_L$ or $\downarrow_L$) of the atom on which the Wannier function is associated, and $U_{Nm\bfk}$ is the projection matrix from the Bloch wave to Wannier orbital. 
%in which the single-particle wave function are expressed using \cite{sasioglu2010wannier}:
%. In this work, we construct the Wannier function from the Bloch wave with the disentangle method \cite{mostofi2014updated} and write it as :
%\begin{eqnarray}
%    |N\bfR\alpha\>=\sum_{m\bfk} e^{i(\bfk\hat{x}-\bfR)}U_{Nm\bfk}^\a|m\bfk\>
%\end{eqnarray}
%where $N$ is the Wannier orbital index, $\bfR$ represents the real-space lattice vector, and $\alpha$ labels the spin index.  On the right-hand side, we use $m$ for the band index and $\bfk$ to denote the wave vector of the Bloch wave function. In the noncollinear DFT framework, the spin dependence is implicitly included in the Bloch states and hence not treated separately. 
\\
\indent
Using Eq.~(\ref{Eq:wan_expand}), we can project the space-time indexed function onto the Wannier basis. After Fourier transformation, the two-particle Green's function becomes:
\begin{eqnarray}
    &&K^{\a\g,\d\b}(\bfq,\w)=
    \sum_{N_1N_2N_3N_4}
    K^{\a\g,\d\b}_{N_1N_3;N_4N_2}(\bfq,\w)
\end{eqnarray}
with
\begin{eqnarray}
&&K^{\a\g,\d\b}_{N_1N_3;N_4N_2}(\bfq,\w)=
    \frac{1}{N_{\bfk}}\sum_{\bfk}\<\bfq|W_{N_1N_2}(\bfk,\bfq)\>
    \nn\\
    &&    ~~~~~~~~~~~~~\times\bar{K}^{\a\g,\d\b}_{N_1N_3;N_4N_2}(\bfk,\bfq,\w)\times
    \<W_{N_3N_4}(\bfk,\bfq)|\bfq\>\nn\\
\end{eqnarray}
where
\begin{eqnarray}
    &&\bar{K}^{\a\g,\d\b}_{N_1N_3;N_4N_2}(\bfk,\bfq,\w)=\sum_{mn}\sum_{\a_L\b_L\g_L\d_L}(f_{m\bfk+\bfq}-f_{n\bfk})
    \nn\\
    &&~~~~
    \times
    \frac{
    U^{\a_L*}_{N_1m\bfk+\bfq}
    U^{\g_L }_{N_3m\bfk+\bfq}
    U^{\d_L*}_{N_4n\bfk}
    U^{\b_L }_{N_2n\bfk}
    }{\w-(\e_{m\bfk+\bfq}-\e_{n\bfk})+i\eta}
    \nn\\
    &&
    ~~~~~\times
    \chi^{\a_L\a,*}_{N_1}
    \chi^{\g_L\g  }_{N_3}
    \chi^{\d_L\d,*}_{N_4}
    \chi^{\b_L\b  }_{N_2}
    \label{Eq:Kab_general}
\end{eqnarray}
where, in $\chi^{\a_L\a}$, the first index, $\a_L=\uparrow_L,\downarrow_L$, is the local spin index while the second index, $\a$, is its component in laboratory coordinate. $|W_{N_1N_2}(\bfk,\bfq)\>$ is the orthonormal transition basis defined as:
\begin{eqnarray}
&&|W_{N_1N_2}(\bfk,\bfq)\>=
|\tilde{\w}_{N_1\bfk+\bfq}\tilde{\w}^{*}_{N_2\bfk}\>
=\nn\\
&&
\sum_{N_3N_4}
\left[\mathcal{O}^{-1/2}(\bfq,\bfk)\right]_{N_1N_2,N_3N_4}
|{\w}_{N_3\bfk+\bfq}{\w}^{*}_{N_4\bfk}\>
\end{eqnarray}
with 
\begin{eqnarray}
    \mathcal{O}_{N_1N_2,N_3N_4}(\bfq,\bfk)=
\<\w_{N_1\bfk+\bfq}\w^{*}_{N_2\bfk}
|\w_{N_3\bfk+\bfq}\w^{*}_{N_4\bfk}\>.
\end{eqnarray}
and has the representation in the spin wave momentum space as:
\begin{eqnarray}
    \<\bfq|W_{N_1N_2}(\bfk,\bfq)\>=
    \int d\bfx e^{i\bfq\cdot\bfx}
    \tilde{\w}_{N_1\bfk+\bfq}(\bfx)\tilde{\w}^{*}_{N_2\bfk}(\bfx).
    \label{Eq:qW}
\end{eqnarray}
\\
\indent
Similarly, in the interaction kernel, the Coulomb interaction becomes:
\begin{eqnarray}
    &&W_{N_1\bfR_1 N_3\bfR_3 N_4\bfR_4N_2\bfR_2}=\int\int
    w^{*}_{N_1\bfR_1}(\bfx)
    w^{}_{N_3\bfR_3}(\bfx)
    \nn\\
    &&~~~~\times W(\bfx,\bfx')
    w^{*}_{N_4\bfR_4}(\bfx')
    w^{ }_{N_2\bfR_2}(\bfx')
    d\bfx d\bfx'.
    \label{Eq:Wnnnn}
\end{eqnarray} 
In the following discussion, we keep the major on-site contribution with $\bfR_1=\bfR_2=\bfR_3=\bfR_4$, such that we can further neglect the momentum dependence in the Coulomb term after Fourier transformation. %Numerically, the Coulomb screening is computed within the Random Phase Approximation in static limit, which requires hundreds bands to converge. To reduce the computational cost, without a full converged value, it's common to introduce a overall rescaling factor \cite{sasioglu2010wannier,skovhus2022magnons}, by:
Numerically, in order to satisfy the Goldstone theorem, we adopt a overall rescaling \cite{sasioglu2010wannier,skovhus2022magnons}:
\begin{eqnarray}
    W\rightarrow \tilde{W}=(1+\lambda) W
    \label{Eq:WlW}
\end{eqnarray}
with a small correction $\lambda$
to set Goldstone mode lying at zero-energy.
%and choose proper value $\lambda$ setting the Goldstone mode to lie at zero-energy. 
Using Eq.~(\ref{Eq:Kab_general}) and Eq.~(\ref{Eq:Wnnnn}), Eq.~(\ref{Eq:BSE}) can be recast as:
\begin{eqnarray}
    &&T^{\a\d,\g\b}_{N_1N_3;N_4N_2}(\bfq,\w)=
    W_{N_1N_3;N_4N_2}\d_{\a\d}\d_{\g\b}+\nn\\
    &&\sum_{\a_1\b_1}\sum_{N_5N_6N_7N_8}W_{N_1N_5;N_6N_2}
    \nn\\
    &&~~~~~~\times K^{\a\a_1,\b_1\b}_{N_5N_7;N_8N_2}(\bfq,\w)
    T^{\a_1\d,\g\b_1}_{N_7N_3;N_4N_8}(\bfq,\w)
    \label{Eq:BSE_wan}
\end{eqnarray}
The reduction of computational complexity now becomes manifest. In a typical DFT calculation, the size of fast Fourier transformation grid is $\sim O(10^5)$, leading to the kernel $T^{\a\g,\d\b}(1,3;4,2)$ with size $\sim O(10^{20})$. In contrast,  the use of  Wannier basis reduces the kernel to $\sim O(10^4)$, making the computation of $T^{\a\g,\d\b}_{N_1N_3;N_4N_2}$ feasible for modern computing platform.
\\
\indent
In principle, the spin-wave spectrum can be derived by specifying $(i,j)=(+,-)$ in Eq.~(\ref{Eq:Rij}) where $\hat{\s}^\pm=\frac{1}{2} ( \hat{\s}^x\pm i \hat{\s}^y)$ are spin-flip operators that convert majority spins to minority spins and vice versa. However, in noncollinear spin systems, the notion of majority and minority spin is no longer well-defined with respect to the global (laboratory) coordinate system. To address this, we utilize the locality of the Wannier functions in Eq.~(\ref{Eq:qW}), which confines spin-flip processes to the same atomic site. This allows us to adopt a local coordinate frame to define the up-spin, down-spin, and the flipping among the two states, which aligns with the  description commonly used in local spin models \cite{akagi2013effect,dos2018spin,dos2020modeling,rezende2020magnon}. As illustrated in Fig.~\ref{Fig:local}, by fixing the $z'$-axis in the local frame, the spin-flip operators retain the same mathematical form, $\hat{\s}^{\pm'}=\frac{1}{2}(\hat{\s}^{x'}\pm i \hat{\s}^{y'})$, but are now defined relative to the local coordinates.
Overall, the dynamical spin susceptibility Eq.~(\ref{Eq:Rij}) becomes:
\begin{eqnarray}
    &&R^{ij,ab}(\bfq,\w)=\nn\\
    &&-
    \sum_{\a\g,\d\b}
    \sum_{N_1N_2\in a}
    \sum_{N_3N_4\in b}
    \sigma^{i^a}_{\b\a}\sigma^{j^b}_{\g\d}
    R^{\a\g,\d\b}_{N_1N_3;N_4N_2}(\bfq,\w)\nn\\
    \label{Eq:Rij_noncol}
\end{eqnarray}
where $a$ and $b$ are the atomic index, $\sigma^{i^a}$ is the Pauli matrix defined in the local coordinate of the $a$th atom, and
\begin{eqnarray}
    &&R^{\a\g,\d\b}_{N_1N_3;N_4N_2}(\bfq,\w)=\frac{1}{N_{\bfk}}\sum_{\bfk}
    \<\bfq|W_{N_1N_2}(\bfk,\bfq)\>\times
    \nn\\
    &&\left[K^{\a\g,\d\b}_{N_1N_3;N_4N_2}(\bfq,\w)+L^{\a\g,\d\b}_{N_1N_3;N_4N_2}(\bfq,\w)\right]
    \<W_{N_3N_4}(\bfk,\bfq)|\bfq\>\nn\\
    \label{Eq:Rij_wLw}
\end{eqnarray}
To compare with the experimental measurement, from the susceptibility function, we can write down the spectral function \cite{skovhus2021dynamic,skovhus2022magnons}:
\begin{eqnarray}
    &&S^{+-}(\bfq,\w)={\rm Tr}_a\left[S^{+-,aa}(\bfq,\w)\right]\nn\\
    &&
    S^{+-,ab}(\bfq,\w)=\frac{1}{2\pi i}\left[R^{+-,ab}(\bfq,\w)-R^{-+,ba}(-\bfq,-\w)\right]\nn\\
    \label{Eq:Sqw}
\end{eqnarray}
It is important to note that the atomic indices $a$ and $b$ in Eq.~(\ref{Eq:Rij_noncol}) and Eq.~(\ref{Eq:Sqw}) represent the intra-atomic spin response: a change in the spin state of atom $b$ induces a corresponding spin flip on atom $a$. In systems containing multiple magnetic atoms within a unit cell, the eigenmodes of magnetic excitations are not limited to isolated single-spin flips. Instead, they manifest as linear combinations of spin fluctuations across multiple atoms,
\begin{eqnarray}
\sigma^{\pm,\lambda}=\sum_{a}C_{\lambda a}\sigma^{\pm^a},
\end{eqnarray}
for which their combination coefficient can be obtained by diagonalizing the $S^{+-,ab}$:
\begin{eqnarray}
S^{+-,ab}(\bfq,\w)=\sum_{\lambda}C^*_{\lambda b}(\bfq,\w) D_\lambda (\bfq,\w)    C_{\lambda a}(\bfq,\w)
\label{Eq:Spm_diag}
\end{eqnarray}
while the sum of contribution from all spin wave mode is equivalent to taking the trace over the atomic index.

\subsection{Ansatz potential method for large-scale magnetic DFT calculations
\label{Subsec:AP-method}}
The starting point of the MBPT calculations are the DFT calculations for noncolinear spin systems, 
where many-body electron problems are mapped onto the single-particle Kohn-Sham equation:
%The DFT calculation formulated under the Kohn-Sham formalism simulates the electronic structure by solving the Kohn-Sham equation: 
\begin{align}
    \sum_{s'}\left\{\left[\frac{-\nabla^2}{2}+V_{\rm KS}(\bfx)\right]\delta_{ss'}+\left[{\boldsymbol{ \sigma}}_{ss'}\cdot \bfB_{\rm KS}(\bfx)\right]\right\}\Psi_{s'}(\bfx)~~~&\nn\\
    =E\Psi_s(\bfx)&
    \label{Eq:KS-noncol}
\end{align} 
%DFT calculation, formulated within the Kohn-Sham framework, has been widely employed over the past few decades to accurately simulate the properties of materials. By introducing an effective exchange-correlation (XC) potential, the Kohn-Sham equations transform the many-body electron problem into a tractable single-particle form: \begin{align}
%    \sum_{s'}\left\{\left[\frac{-\nabla^2}{2}+V_{\rm KS}(\bfx)\right]\delta_{ss'}+\left[{\boldsymbol{ \sigma}}_{ss'}\cdot \bfB_{\rm KS}(\bfx)\right]\right\}\Psi_{s'}(\bfx)~~~&\nn\\=E\Psi_s(\bfx)& \label{Eq:KS-noncol}\end{align} 
where $V_{\rm KS}$ and $\bfB_{\rm KS}$ encode the many-body interaction by a scalar field and a local effective magnetic field, respectively. These potentials are determined through a self-consistent procedure, involving iterative updates of the charge and magnetization densities until convergence is achieved. However, the computational cost of standard DFT scales as  $O(N_{\rm e}^3)$ with respect to the number of electrons $N_{\rm e}$, which significantly limits its applicability to large-scale magnetic structures such as spin helices and skyrmions. 
\\
\indent
To overcome this limitation, Chen $et~al.$  \cite{chen2025topological} recently proposed an efficient modeling approach that determines the Kohn-Sham potential without resorting to computationally expensive self-consistent calculations. Drawing inspiration from the double-exchange model  \cite{anderson1955considerations}, the magnetic part of Kohn-Sham potential in a unit cell located at site-$r$ is expressed as a functional of the local spin moment:
\begin{eqnarray}
    &&{\bfB}^r_{\rm KS}(\bfx)={\bfB}_{S}[\{\bfS\}_r](\bfx)
    \label{Eq:Bex_S}
\end{eqnarray}
where $\{\bfS\}_r$ denotes the set of local spin moments in the vicinity of site-$r$, typically within a cutoff radius of one or two lattice constants from the center of the unit cell. This potential is then constructed via an ansatz relation that explicitly expresses its dependence on the surrounding spin configuration:
\begin{eqnarray}
    &&{\rm B}^i_{S}[\{\bfS\}_r](\bfx)=\sum_{a,j} J^i_{a,j}(\bfx)s^j_a
    \nn\\
    &&~~~~
    +\sum_{ab,jk} J^i_{ab,jk}(\bfx)s^j_as^k_b+O(s^{3})+\cdots
    \label{Eq:B_JS}
\end{eqnarray}
where $i,j,k,\cdots=x,y,z$ are the direction indices. ${\bf s}_a \in \{\bfS\}_r$ is the nearby spin moment denoted by index $a$, which is defined as the average of magnetization:
\begin{eqnarray}
    {\bf s}_a=\int_{\bfx \in R_a} \bfm(\bfx)d\bfx 
    \label{Eq:local_s}
\end{eqnarray}
where $R_a$ is a cutoff radius around the position of $a$th nearby atom. The coefficient $J$'s are further determined by fitting the ${\bf B}_{\rm KS}$ data obtained from DFT calculations on a supercell containing spins of $\{\bfS\}_r$ in Eq.~(\ref{Eq:Bex_S}). Given that the spin dependence in Eq.~(\ref{Eq:Bex_S}) decays rapidly with the distance between the spin site and the target cell, it is not necessary to perform full-scale calculations for the entire noncollinear spin structure. Instead, one can employ relatively small supercells that include only the relevant neighboring spins to solve the Kohn-Sham equations. This approach significantly reduces the computational cost and makes the calculation feasible for complex spin textures. Finally, the potential of the full magnetic structure can be obtained by the summation over all local contribution:
\begin{eqnarray}
    \bfB_{\rm KS}(\bfx)=\sum_{r}{\bfB}_{S}[\{\bfS\}_r](\bfx)
    \label{Eq:BKS_sum}
\end{eqnarray}
while the scalar potential $V_{\rm KS}$ is approximately independent from the spin structure. Overall, by substituting Eq.~(\ref{Eq:BKS_sum}) into the Kohn-Sham equation Eq.~(\ref{Eq:KS-noncol}), we can solve for the Kohn-Sham orbitals and the corresponding energy spectrum. 
\\
\indent
Besides directly applying the constructed potential, Eq.~(\ref{Eq:BKS_sum}) to the Kohn-Sham equation, this result is also compatible with the Wannier tight-binding model. From a non-magnetic Hamiltonian, $H_{\rm NM}$ and its hopping integral:
\begin{eqnarray}
    H^{\text{non-mag}}_{NM}(\bfR)=\<w_N {\bf 0}|H^{\text{non-mag}}|w_M\bfR  \>
\end{eqnarray}
where $|w_{M}\bfR \>$ is the $M$th Wannier function localizing at site-$\bfR$, the hopping element in the tight-binding Hamiltonian of the magnetic system can be constructed as:
\begin{eqnarray}
    H^{\text{non-mag}}_{NM}(\bfR)=H^{\text{non-mag}}_{NM}(\bfR)+\<w_N{\bf 0}|\bfB_{\rm KS}(\bfx)|w_M\bfR \>.\nn\\
    \label{Eq:Hsp}
\end{eqnarray}
The system now holds a different translational symmetry from the crystal. Instead, it follows the periodicity of the spin pattern. Consequently, we need to use the supercell which involves one period of spin texture to re-define the unit cell to apply the Bloch theorem and Fourier transform the tight-binding Hamiltonian to $\bfk$-space, which can give us the required eigenenergy, $\e_{m\bfk}$, and projection matrix, $U_{Nm\bfk}$, to construct the two-particle Green's function, Eq.~(\ref{Eq:Kab}) to compute the magnetic response in noncollinear spin system. Without digression from the main idea, we leave the detailed discussion to Appendix~\ref{sect:unit2super}.

\section{Magnetic excitation in noncollinear spin helix in $\bf LiCu_2O_2$
\label{sect:application}}
\begin{figure}[t]
    \centering   \includegraphics[width=0.9\linewidth]{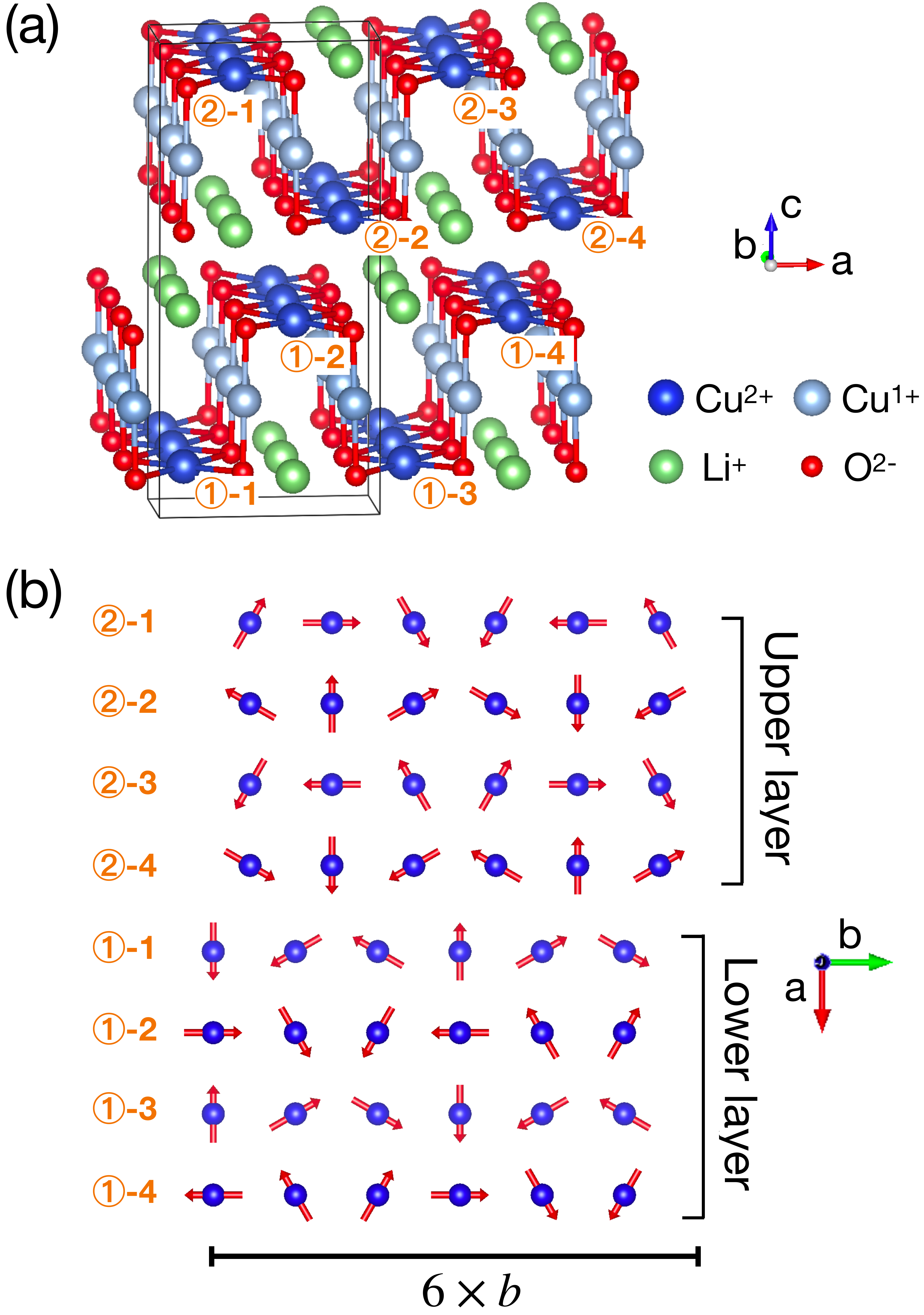}
    \caption{(a) LiCu$_2$O$_2$ crystal structure and atomic compositions. The magnetic structure is mainly formed by the net spin momentum on $\rm Cu^{2+}~(0.52\mu_B)$  and on $\rm O^{2-}~(0.18\mu_B)$ as spin helix chains along $b$-axis. Nearby chains are connected by non-magnetic Cu$^{1+}$ ions, forming a layered structure.  (b) Schematic of the Cu$^{2+}$ spin orientation from the top view along $c$-axis, while, for a clear illustration, the finite magnetic moments on O$^{2-}$ ions are hidden. Each layer (\textcircled{1} and \textcircled{2}) contains four spin-helix chains rotating within the $ab$-plane and propagating along the $b$-axis. In this work, a commensurate magnetic pitch of 6 unit cells is adopted. Experimental measurements indicate a $\pi$-phase shift between chain-1 and chain-3, as well as between chain-2 and chain-4, while a $\pi/2$-phase shift between chain-1 and chain-2, as well as between chain-3 and chain-4 are expected. Between layer \textcircled{1} and layer \textcircled{2}, there is a further $5\pi/6$-phase shift in relative spin orientation.  }
    \label{Fig:LiCu2O2}
\end{figure}

$\rm LiCu_2O_2$, a type-II multiferroic, is a prototypical example of a quasi-one-dimensional frustrated antiferromagnet. 
%It contains both nonmagnetic Cu$^+$ and magnetic Cu$^{2+}$ ions, forming $S=1/2$ spin chain along the $b$ axis of the orthorhombic crystal. 
It consists of Li$^+$, Cu$^{2+}$, Cu$^{1+}$, and O$^{2-}$ ions, with valences given as nominal values. 
While Cu$^{1+}$ ions are nonmagnetic, magnetic Cu$^{2+}$ ions form an S=1/2 spin chain along the $b$ axis of the orthorhombic crystal. Besides, O$^{2-}$ ions with finite magnetic moments bridge neighboring Cu$^{2+}$ ions along the chain (see Fig.~\ref{Fig:LiCu2O2}) with an angle $\angle_{\rm Cu-O-Cu}\approx 94^\circ$ \cite{hibble1990licu2o2,berger1991note,berger1991structure}. Neutron scattering and NMR experiments have revealed an incommensurate spin-spiral magnetic structure with wave vector $\bfq=(0.5,0.827,0.0)$, where the Cu$^{2+}$ spins rotate within a plane below the critical temperature $T_{\rm c}=24$ K \cite{masuda2004competition,masuda2005spin,park2007ferroelectricity,bush2018exotic}.
\\
\indent
Despite the consistent experimental observation of the spin-helix state, its microscopic origin remains under debate. Some theoretical studies highlight the role of the Dzyaloshinskii–Moriya (DM) interaction, which becomes active due to local inversion symmetry breaking \cite{furukawa2010chiral}. Others instead attribute the magnetic ordering to the superexchange interaction between Cu$^{2+}$ ions \cite{gippius2004nmr,bush2018exotic,Boidi2023determination}. However, most existing studies rely on localized spin models with parameters obtained either empirically or from first-principles calculations, and primarily focus on the Cu$^{2+}$ ion while overlooking the nontrivial spin moment on O$^{2-}$ ion.
\\
\indent
In this section, we apply our formalism to $\rm LiCu_2O_2$ to predict magnetic instability and the corresponding excitation spectrum alongside revealing the underlying physical mechanism. We begin by introducing the construction of the Wannier basis and demonstrating how to accommodate the non-magnetic tight-binding model with the magnetic exchange field derived from the ansatz potential method. To predict the magnetic stable state, we first analyze the magnetic response in the long-wavelength limit ($\bfq=\Gamma$), highlighting the impact of spin noncollinearity on the spectral function. The results indicate a magnetic instability extending beyond the unit cell. Motivated by this finding, we proceed to compute the spin-wave dispersion to probe finite-$\bfq$ effects. Our analysis reveals a dominant unstable magnon mode emerging at $\bfq=(0.5,q_y,0.0)$, where the value of $q_y$ is sensitive to the strength of the magnetic exchange interaction. We further perform a detailed numerical analysis to quantify this dependence. In parallel, we investigate the role of magnetic moment on the ligand $\rm O^{2-}$ ions in effective ferromagnetic interactions between primary magnetic $\rm Cu^{2+}$ sites, underscoring their influence on the magnetic excitation. Finally, by adopting appropriate magnetic exchange strength, we present our prediction of the magnon excitation spectrum in the spin helix state and compare it with previous experimental studies.
%In this section, we will apply the formalism introduced in Sect.~\ref{Sec:Theory} to investigate the noncollinear spin ground state the corresponding excitation structure from first principles, where the spin moment on Cu$^{2+}$ ion and O$^{2-}$ ion are treated under equal footing. We will first present the result of DFT calculations with the construction of Wannier orbitals and discuss how ansatz potential method works in this material. Based on the result, we investigate the magnetic response in the long-wavelength limit ($\bfq=\Gamma$), revealing \blue{[YN: should be "revealing"?]} \green{[HYC: corrected]} the effect of spin-noncolinearity on the spectrum function. Furthermore, using the ferromagnetic ground state, we compute the magnon dispersion from which we predict the pitch of the spin helix structure and discussed its relation to exchange couplings. Before closing, we present our calculation of the magnon dispersion in the spin spiral ground state and compare with the experimental measurement.
\subsection{Electronic structure, Wannier tight-binding model, and ansatz potential}
\begin{figure}
    \centering
    \includegraphics[width=1.0\linewidth]{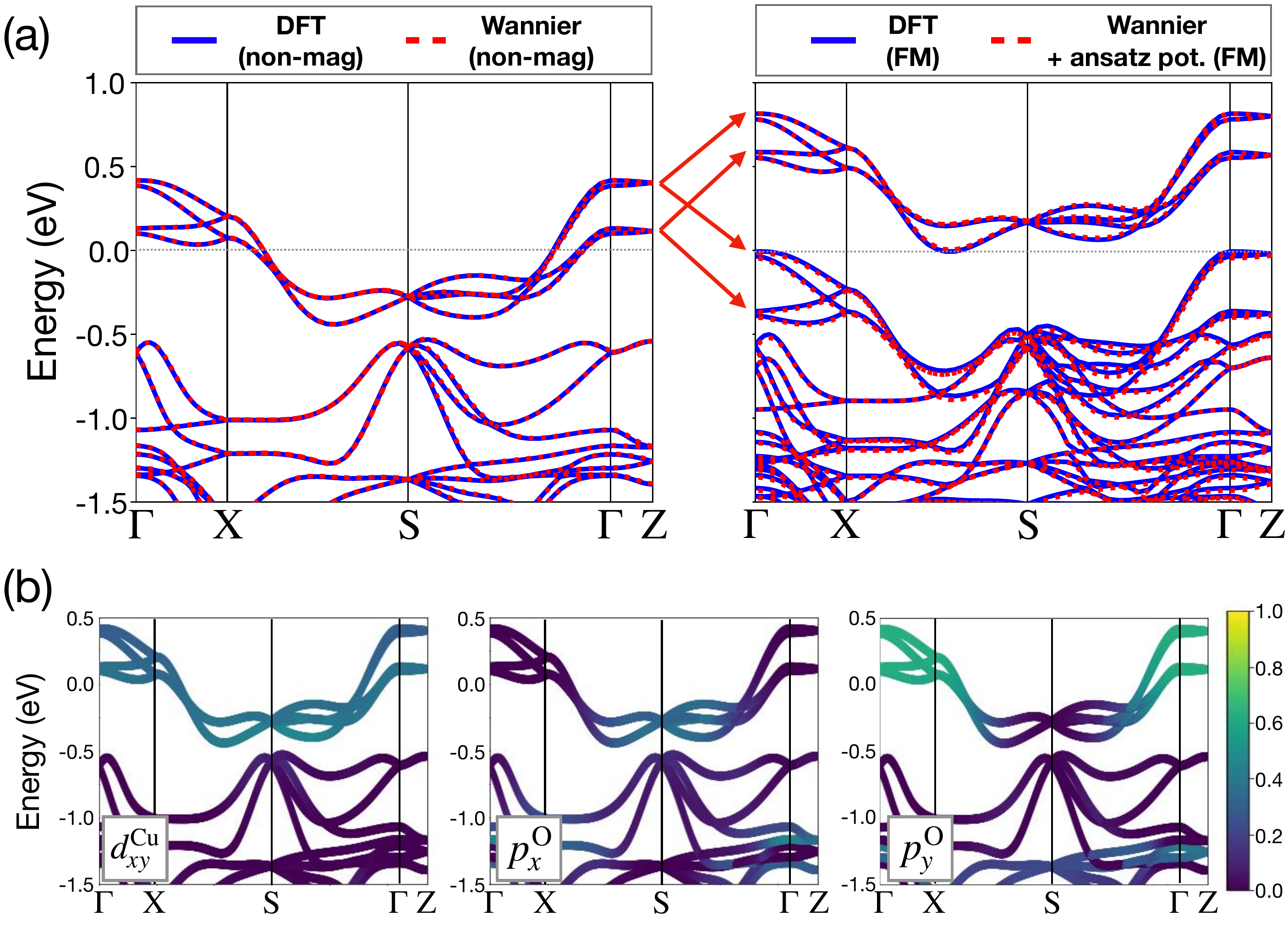}
    \caption{(a) Left: Band structure of $\rm LiCu_2O_2$ in non-magnetic ground state computed directly in DFT and in Wannier tight-binding model. Right: Band structure of $\rm LiCu_2O_2$ in ferromagnetic (FM) ground state computed directly in DFT (Blue line). In addition, we combine the non-magnetic Wannier tight-binding model with the magnetic potential obtained from the ansatz potential method. The spin-splitting about $1.0$ eV is correctly reproduced with an error less than 20 meV. (b) Band projection on the $d_{xy}$ orbital of $\rm Cu^{2+}$ ions and the $p_{x,y}$ orbital of $\rm O^{2-}$ ions. }
    \label{Fig:band}
\end{figure}
We carry out the DFT calculation using {\sc Quantum espresso} package with pseudopotential obtained under the general gradient approximation (GGA). The computational details are summarized in the Appendix \ref{Append:detail}. We start with the ferromagnetic structure and obtain magnetic moment $\mu_{\rm Cu^{2+}}=0.52 \mu_B$ and $\mu_{\rm O^{2-}}=0.18 \mu_B$ consistent with early reports \cite{koc2025structural,zatsepin1998valence}. In order to construct the spin-symmetric Wannier orbtals and applying the ansatz potential method, we remove the magnetic part of the exchange potential and conduct the non-self-consistent field (nscf) calculation to obtain the nonmagnetic Kohn-Sham orbital. The corresponding electronic band structure is presented in Fig.~\ref{Fig:band}(a). Without the magnetic field, we construct Wannier tight-binding model using $d$-orbitals of both $\rm Cu^{2+}$ and $\rm Cu^{1+}$ ions and $p$-orbitals of $\rm O^{2-}$ ions, obtaining results without visible error. On the other hand, we generate a ferromagnetic exchange potential using the ansatz potential method and add its effect into the nonmagnetic tight binding model. We summarize the details of ansatz potential in Appendix \ref{Append:AP-method}.  
The result shows high agreement with the direct DFT calculation of the ferromagnetic state. The exchange splitting about 1.0 eV is correctly reproduced with minor deviation less than 20 meV in the band structure. As emphasized in Fig.~\ref{Fig:band}(a), the exchange splitting breaks the degeneracy of the four entangled bands near the Fermi level, transforming the ferromagnet from a metal to semimetal with a tiny negative gap of about 10 meV. By decomposing the wave function, as shown in Fig.~\ref{Fig:band}(b), the split bands mainly consists of the $d_{xy}$ orbitals of Cu$^{2+}$ ion with components from the $p_{x,y}$ orbitals of O$^{2-}$ ions. This results clearly explain the source of the magnetic moments of Cu$^{2+}$ ions and O$^{2-}$ ions. Consequently, in the following when discussing the magnetic excitation, we will focus on the spin-flipping process on these three kinds of orbitals, since other orbitals are buried under the Fermi level which merely contributes according to Eq.~(\ref{Eq:Kab_general}).

\subsection{ Ferromagnetic instability and noncollinear effect at $\bfq=\Gamma$ }
\begin{figure}[t]
    \centering
    \includegraphics[width=0.95\linewidth]{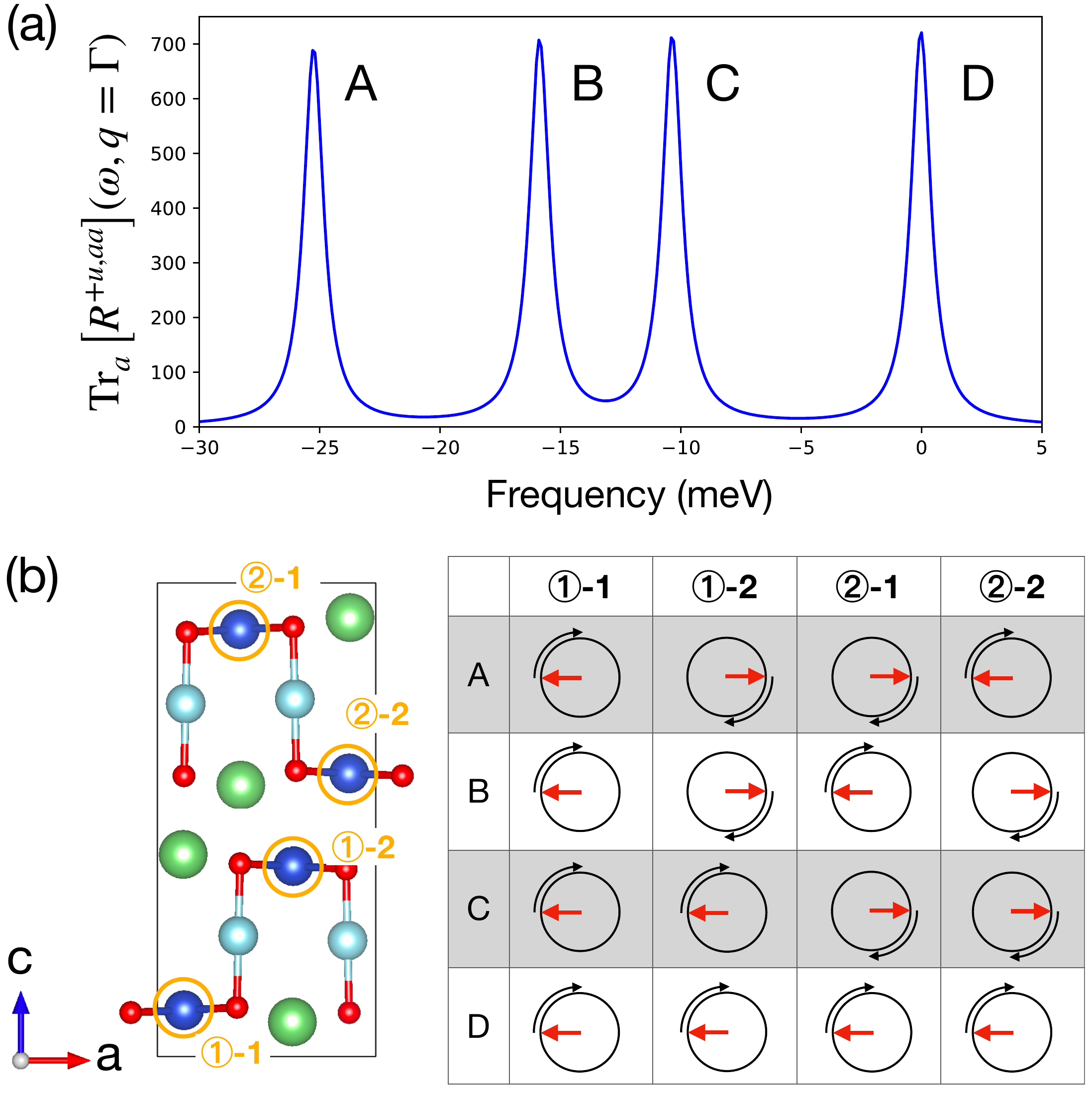}
    \caption{(a) Zero-momentum magnetic response function. Four distinct peaks correspond to the four principal spins of the Cu$^{2+}$ ions in one unit cell. D-peak represents a co-moving, in-phase rotation mode corresponding to the Goldstone mode arising from spontaneous symmetry breaking, and thus incurs no excitation energy. In contrast, the other three peaks lie in the negative energy range, indicating magnetic instability.
    (b)  Spin precession patterns viewed along the rotation axis of each spin, as indexed in the left panel. The patterns are obtained by Eq.~(\ref{Eq:Spm_diag}) at the respective peak positions.}
    \label{Fig:Gamma}
\end{figure}
\begin{figure}[t]
    \centering
    \includegraphics[width=0.8\linewidth]{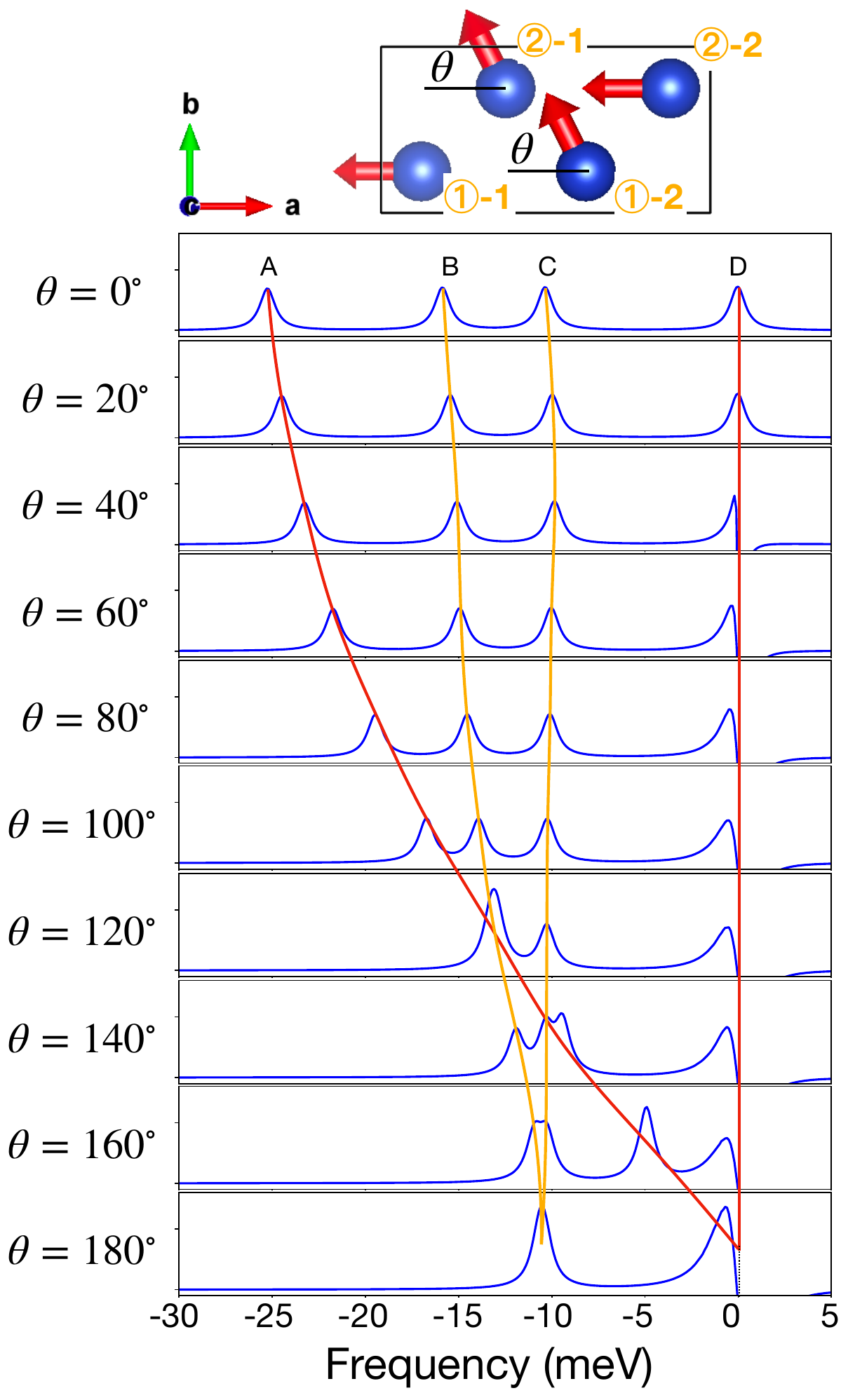}
    \caption{The magnetic response of different magnetic ground state illustrated in the top panel. The antiferrormagnetic force between \textcircled{1}-1,\textcircled{1}-2 spins and \textcircled{2}-1,\textcircled{2}-2 spins is relaxed with the increasing of relative angle $\theta$. Thus, the instability of A-peak is reduced with its energy raised. While sharing the similar intra-layer antiferromagnetic interaction,  the energy position of B-peak also slightly increases. On the contrary, C-peak remains. The positive part of D-peak structure is broken and acquire a negative value, which is a feature of antiferromagnetic system \cite{skovhus2022magnons}.
    }
    \label{Fig:rot}
\end{figure}
To analyze the ground state magnetic structure, we first focus on the four primary spins on the Cu$^{2+}$ ions within a unit cell \footnote{According to the DFT calculation on one unit cell, the secondary spins on O$^{2-}$ always lie along with the magnetic moment on the nearest Cu$^{2+}$ ions parallelly.} and investigate the corresponding magnetic response in the long wavelength limit. We start with the collinear ferromagnetic state and compute the magnetic response spectrum as in Fig.~\ref{Fig:Gamma}(a) for which we obtain four peaks within 30 meV separation. To determine the corresponding spin excitation, for each peak, we apply Eq.~(\ref{Eq:Spm_diag}) to extract the eigenmode of the magnon state and collect the results in Fig.~\ref{Fig:Gamma}(b). The result shows that the mode with highest energy (D-peak) is a co-moving precession motion, in which the four spins rotate without relative phase delay. Since the spin-orbit coupling is not included in our calculation, this acoustic co-moving motion should be identified as the  Goldstone mode of ferromagnetic symmetry breaking, costing no energy to excite. After setting Goldstone mode to zero energy level by tuning the Coulomb interaction as in Eq.~(\ref{Eq:WlW}), we found the other three peaks remain in the negative energy range, which indicates the instability of ferromagnetic structure and reveals the nature of antiferromagnetic interaction among the spins. For example, in A-peak, intra-layer spins \textcircled{1}-1,\textcircled{1}-2 and \textcircled{2}-1,\textcircled{2}-2 tend to pointing to opposite direction, suggesting as relative stable antiferromagnetic structure $(S_{ \textcircled{1}\text{-}1},S_{\textcircled{1}\text{-}2},S_{\textcircled{2}\text{-}1},S_{\textcircled{2}\text{-}2})=(\uparrow,\downarrow,\downarrow,\uparrow)$. This conjecture is confirmed in collinear DFT energy calculation, where we obtained the energy hierarchy: 
$   E_{\uparrow\downarrow\downarrow\uparrow} < E_{\uparrow\downarrow\uparrow\downarrow} <
    E_{\uparrow\uparrow\downarrow\downarrow} <
    E_{\uparrow\uparrow\uparrow\uparrow}
$, for the four configuration corresponding to the four rotational modes.
\\
\indent
To explicitly illustrate the relation between the negative energy mode and the antiferromagnetic structure, 
we change the ground state by varying the relative angle between \textcircled{1}-1,\textcircled{2}-2 and \textcircled{1}-2,\textcircled{2}-1 spins and compute the magnetic response as in Fig.~\ref{Fig:rot}. The results show that as the angle $\theta$ increases, the energy of peak-A gradually rises, i.e. becoming less unstable, and eventually merges into the acoustic magnon mode when the system reaches the fully-antiferromagnetic state at $\theta=180^{\circ}$. Besides, the energy of peak-B also slightly increases since it shares the same major antiferromagnetic pattern with peak-A between \textcircled{1}-1 and \textcircled{1}-2, as well as \textcircled{2}-1,\textcircled{2}-2 spins, which are not favored by peak-C. The remaining negative-energy modes associated with peak-B and peak-C also indicate that the magnetic structure cannot be stabilized within a single unit cell, calling for an analysis on the finite-momentum excitation.
%Additionally, the energy of peak-B also slightly increase, while peak-C remains unchanged. These observation suggest that the peak-B is primarily induced by the antiferromagnetic force between $(\textcircled{1}$-$1 \leftrightarrow \textcircled{2}$-$1)$ and $(\textcircled{1}$-$2 \leftrightarrow \textcircled{2}$-$2)$, whereas the peak-C arises from the interaction between $(\textcircled{1}-1 \leftrightarrow \textcircled{2}$-$2)$ and $(\textcircled{1}$-$2 \leftrightarrow \textcircled{2}$-$1)$. This distinction is due to the fact that the spins of \textcircled{1}-1(\textcircled{1}-2) and \textcircled{2}-2(\textcircled{2}-1) remain parallel as $\theta$ increases. With the nature of antiferromagnetic interactions and their associated magnetic excitations clarified, we conclude that the magnetic structure cannot be stabilized merely by rotating the the four spins within a unit cell. Rather, the stable noncollinear spin configuration is expected to  extend across multiple unit cells, and a further discussion on finite-$\bfq$ instability is required. 
\\
\indent
Before closing, we put a note on the peak shape of the acoustic mode. It can be observed that the peak shape becomes asymmetric and acquires a negative component when $\theta>40^\circ$. We attribute the effect to the time reversal process which is forbidden in the ferromagnetic state but partially appears when there is antiferromagnetic component. In the fully time reversal symmetric case, $\theta=180^\circ$, the negative part should cancel out the positive part, so the acoustic peak disappears \cite{skovhus2022magnons}. Here, for visual clarity, we keep the peak finite by slightly increasing the $\lambda$ parameter in the Coulomb interaction, which could introduce minor error to other peaks but does not affect the above discussion.

\subsection{Analyzing magnetic instability from magnon dispersion and dependence on the exchange splitting \label{sect:mag_disp} }
We further extend the discussion of magnetic instability to finite momentum by investigating the energy dispersion. Compared to the long wavelength limit, determining the finite momentum spectrum is more delicate due to the problem that the magnon dispersion is highly sensitive to both the electron hopping amplitude and the exchange splitting. In this context, conventional DFT is known being insufficient to capture the local electron correlation, often leading to deviations in the predicted spin exchange splitting. This limitation has been shown to hinder first-principles simulations of spin waves, prototypical systems such as ferromagnetic Ni \cite{sasioglu2010wannier} and antiferromagnet Cr \cite{skovhus2022magnons}, and typically requires additional corrections to achieve agreement with experimental observations. Similarly, determining the exchange splitting from first-principle in $\rm LiCu_2O_2$ remains an open question and is reflected in the variability of the computed band gap structure. Early studies show that when using the LDA or GGA for exchange effects, the resulting band structures exhibit indirect transitions with vanishing or even negative gap energies. In contrast, applying hybrid functionals such as mBJLDA or incorporating on-site Coulomb interactions via the DFT+U method leads to significantly larger band gaps—approximately 0.931 eV and 0.66 eV, respectively. Furthermore, a previous study \cite{moser2017electronic} reported that even dynamical mean-field theory (DMFT) fails to fully capture correlation effects in $\rm LiCu_2O_2$. 
\begin{figure}[t]
    \centering
    \includegraphics[width=0.95\linewidth]{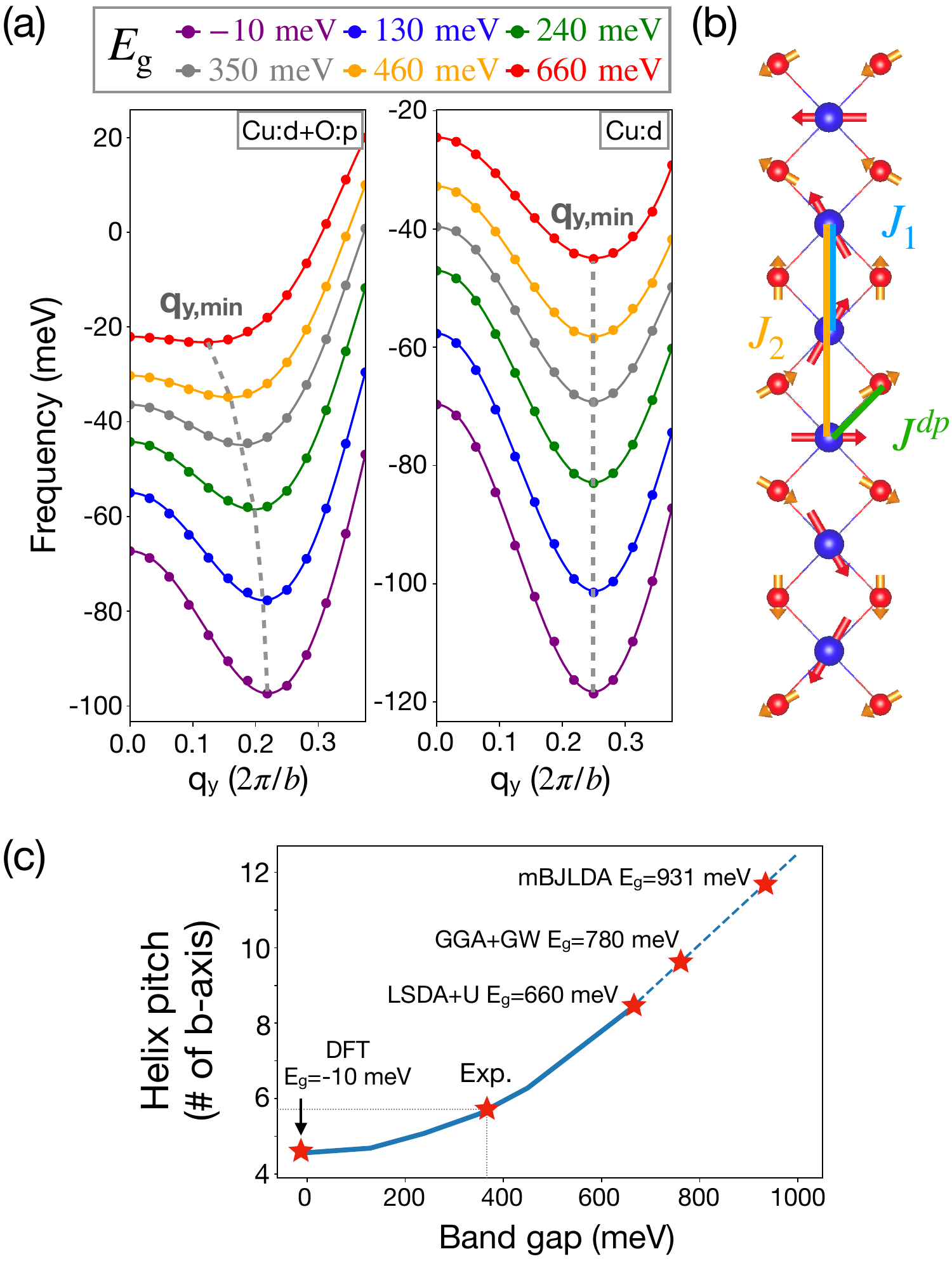}
    \caption{(a) Magnon dispersion of the lowest energy mode along $\bfq=(0.5,q_y,0.0)$ path. We compute with different spin exchange splitting which corresponds to different band gaps $E_{\rm g}$. Besides, two calculations involving different orbitals in the scattering channel is conducted, (left) using $d_{xy}$ on Cu$^{2+}$ and $p_{x,y}$ on O$^{2-}$ ions, (right) using $d_{xy}$ on Cu$^{2+}$ only.   It shows that when we consider orbitals from both ions,  the minimum energy point tends to having smaller momentum $q_{y,\rm min}$ as the exchange force raises. If only the Cu$^{2+}$ orbitals are involved, $q_{y,\rm min}$ remains a constant. 
    (b) Top view from $c$-axis for the \textcircled{1}-1 chain with spins on both Cu$^{2+}$ ions and O$^{2-}$ ions. In the perspective of the spin model, the inter-$\rm Cu^{2+}$ couplings are denoted as $J_1,~ J_2$ while the $\rm Cu^{2+}$-$\rm O^{2-}$ coupling is denoted as $J^{dp}$. 
    (c) We summarize the relation between the  band gap energy and the corresponding pitch of spin helix along y-direction $(=1/q_{y,\rm min})$. The star symbol is applied to indicate the gap value numerically obtained under different condition, including: pure DFT, LSDA+U \cite{zatsepin1998valence}, GGA$+$GW correction, adopting the mBJLDA hybrid function \cite{koc2025structural}, and also the expected value which is expected to reproduce the experimental pitch (denoted by EXP.).}
    \label{Fig:EXvsDISP}
\end{figure}
\\
\indent
To investigate the effect of exchange splitting on magnetic excitations, we compute the magnon dispersion of the lowest-energy mode along the $\bfq = (0.5, q_y, 0.0)$ path \footnote{Based on our calculation, the overall dispersion along $\bfq = (0.0, q_y, 0.0)$ has energy higher than the one along $\bfq = (0.5, q_y, 0.0)$ by $\sim$ 30 meV.} for several values of spin exchange splitting, and present the results alongside the corresponding band gaps in Fig.~\ref{Fig:EXvsDISP}(a). Specifically, we rescale the on-site matrix element $\<w_N{\bf 0}|\bfB_{\rm KS}(\bfx)|w_M\bfR \>$ for $\bfR={\bf 0}$ where $w_N$ and $w_M$ denote the Wannier orbitals associated with Cu$^{2+}$ ions in the tight-binding Hamiltonian, Eq.~(\ref{Eq:Hsp}). 
The common feature across all energy dispersions shows that the ferromagnetic structure becomes more unstable against magnetic excitations as they acquire finite momentum. Starting from the $\Gamma$-point, the excitation energy decreases and reaches a minimum around $\bfq_{y,{\rm min}}=0.15\sim 0.22$ which is in close agreement with the experimental value $\bfq_{\rm exp.}=(0.5, 1-0.173, 0.0)$. This negative mode at finite momentum suggests that the magnetic instability extends beyond a single unit cell—that is, the stable magnetic ground state adopts a long-period spin structure with a pitch of $1/q_{y,{\rm min}}$ along the $y$-directions. Furthermore, we highlight the trend that the minimal momentum $q_{y,{\rm min}}$ decreases with the enlargement of the band gap, indicating a reduction in antiferromagnetic interaction, which prolongs the pitch of stable spin configuration, as the exchange splitting enhanced.
%These two behaviors feature the reducing of antiferromagnetic interaction among the spins, which is consistent with the conventional estimation of the spin interactions, where $J\propto t^2/U$, with $t$ being the hopping amplitude and $U$ is the onsite interaction proportional to the spin splitting. 
\\
\indent
To analyze the dependence, we remove the contribution of $p_{x,y}$-orbitals on the $\rm O^{2-}$ ions from the interacting kernel, Eq.~(\ref{Eq:BSE_wan}), which effectively turn off the coupling between the spins on Cu$^{2+}$ ions and spins on $\rm O^{2-}$ ions, denoted by $J^{dp}$ in Fig.~\ref{Fig:EXvsDISP}(b). As presented in the right panel of Fig.~\ref{Fig:EXvsDISP}(a), the dependence disappears and the location of dispersion minimum becomes a constant at  $q_{y,{\rm  min}}=0.25$. This result suggests that the nearest-neighbor coupling $J_1$ is considerably small compared to the next-nearest-neighbor antiferromagentic coupling $J_2$ since, in the 1D-spin chain, the $q_{\rm min}$ can be estimated by
\begin{equation}
 q_{\rm min}=\frac{1}{2\pi}\arccos(\frac{J_1}{4J_2}) \rightarrow 0.25~~{\rm as}~~ 4J_2 \gg J_1.   
\end{equation}
The origin of $J_1$ is discussed in an early study \cite{Boidi2023determination} in term of superexchange interactions; the Kanamori-Goodenough Cu $\leftrightarrow$ O $\leftrightarrow$ Cu superexchange path give raise to a ferromagnetic interaction and an antiferromagnetic interaction from the Cu $\leftrightarrow$ O $\leftrightarrow$ O $\leftrightarrow$ Cu path. Besides, our tight-binding Hamiltonian also suggests nontrivial direct Cu-Cu exchange. Here, we attribute the smallness of $J_1$ to the cancellation among these contributions. 
\\
\indent
In additional to the weak $J_1$ interaction, as we restore the $\rm O^{2-}$ ions effect, the finite magnetic dipole moment on $\rm O^{2-}$ can also contribute an effective ferromagnetic coupling $J^{\rm eff.}\propto (J^{dp})^2$  when we constructed an one-site effective spin model by integrating out the $\rm O^{2-}$ degree of freedom, which becomes the dominant coupling among nearest-neighbor $\rm Cu^{2+}$ spins  in this system. In this context, as we enhance the on-site exchange spin splitting, only the next-nearest-neighbor antiferromagnetic $J_2$ coupling is suppressed by $J_2\sim t^2_2/U$ while the nearest-neighbor $J^{\rm eff.}$ remains unchanged. Consequently, the increase of $U$ raises the ratio $J^{\rm eff.}/J_2$, thus resulting in the reduction of $q_{\rm min}$, which explains the behavior observed in the left panel of Fig.~\ref{Fig:EXvsDISP}(a).
\\
\indent
Last, we present the relation between the band gap and the computed pitch of the spin helix in Fig.~\ref{Fig:EXvsDISP}(c), as we vary the exchange splitting. In the figure, we highlight representative band gap values from both our calculations and previous studies, including band gaps obtained using standard DFT, LSDA+U, GGA+GW, and the mBJLDA hybrid functional \cite{zatsepin1998valence,koc2025structural}. By extrapolating the relation between the band gap and the predicted magnetic pitch, we find that methods beyond DFT yield longer periodicities, whereas the standard DFT calculation produces a shorter pitch as it in general underestimates the exchange splitting. Among them, the experimental pitch ($\approx 5.78$) resides in the intermediate region, calling for more sophisticate approach to compute the exchange splitting. 
\subsection{Magnon dispersion in helimagnetic phase}
\begin{figure}[t]
    \centering
    \includegraphics[width=0.8\linewidth]{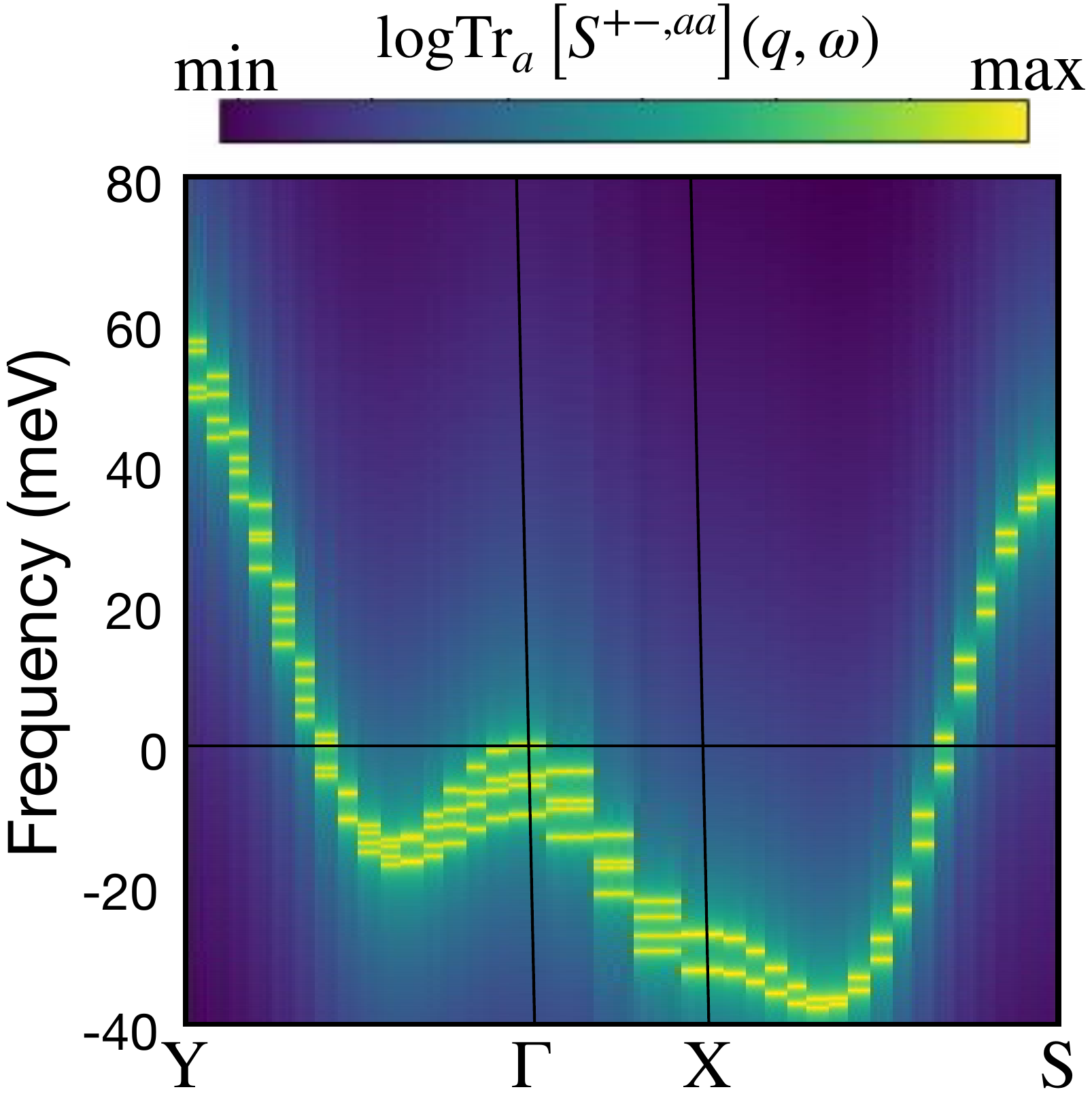}
    \caption{Magnetic response spectrum of ferromagnetic state along symmetry path: Y-$\Gamma$-X-S. Two minima appear with finite momentum, indicating the magnetic instability over multiple unit cell, while the instability along the $x$-direction further reduce the energy along X-S line making $\bfq=(0.5,0.166,0.0)$ the global minimum in consistent with experiment.} 
    \label{Fig:FM_disp}
\end{figure}
\begin{figure}[t]
    \centering
    \includegraphics[width=0.96\linewidth]{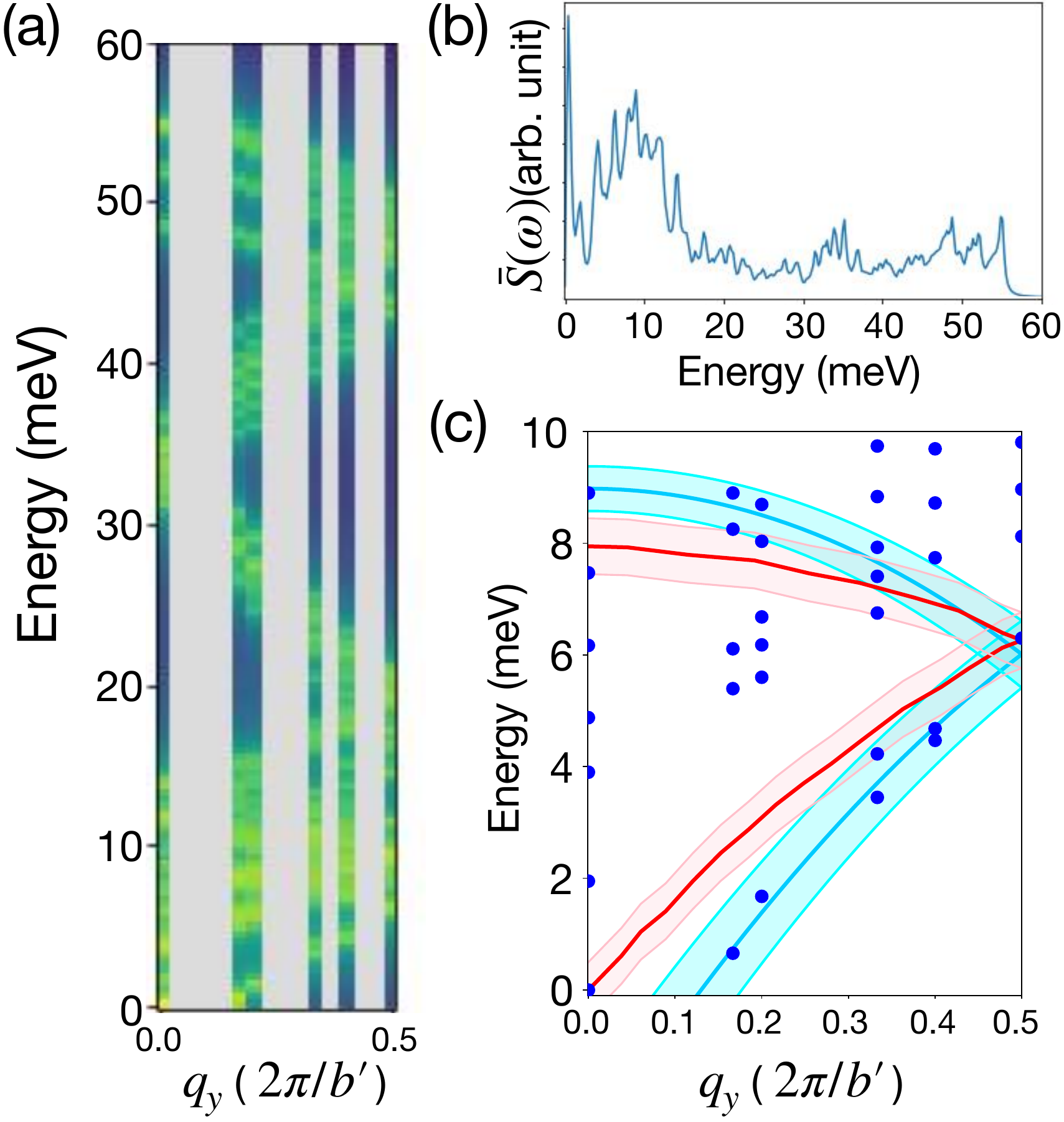}
    \caption{(a) Magnetic response spectrum of the $2\times 6\times 1$ helimagnetic phase along the $y$-direction. The spectral function at each momentum. The spectral function of each momentum $\bfq$ is computed independently. Data are compiled from two separate calculations using $4\times 5\times 6$ and $4\times 6\times 6$ $\bfk$-point grids in the folded Brillouin zone. Overall, the spectrum reveals a set of strongly dispersive magnon states extending from $\omega=0$ to $60$ meV, as well as a set of weakly dispersive low-energy modes confined within $\omega=0$-$15$ meV. (b) Momentum-averaged spectrum along the $y$-direction, serving as an indicator of the magnon density of states. A distinct lump appears near $\omega=10$ meV, suggesting the presence of a band edge or a possible gapped optical mode within this energy range. (c) Comparison between the computed peak positions (blue dots), a fitted dispersion curve for the lowest-energy mode (light blue line with shaded error region), and experimental data from Ref.~\cite{masuda2005spin} (red line with pink error band). The shaded regions represent the uncertainty in both the least-squares fitting and experimental measurements. The remaining negative energy states a finite momentum corresponds to a long pitch of approximate 60 $b$-lattice constants, which can be regarded as numerical artifacts. In contrast, the fitted dispersion shows excellent agreement with the experimental results, with a deviation of less than 1.5 meV.      }
    \label{Fig:261}
\end{figure}
To process to simulate the spin excitation in the spin helix, we interpolate the trend in Fig.~\ref{Fig:EXvsDISP}(c) and set the exchange splitting that reproduces a pitch of six---the commensurate value closest to the experimental result---for the following analysis. We first present the corresponding magnetic spectrum of FM state in Fig.~\ref{Fig:FM_disp}, along high-symmetry lines. The spectrum exhibits all the features discussed above: four peaks at $\Gamma$-point and a global minimum at $\bfq_{\rm min}=(0.5, 1/6,0.0)$. Additionally, a local minimum appears midway along the $\rm Y-\Gamma$ line and $\rm X-S$ line while the later one lowers approximately by 20-30 meV. %Overall, the full magnon bandwidth spans approximately 100 meV.
\\
\indent
We further apply the spin-helix configuration shown in Fig.~\ref{Fig:LiCu2O2}(b) as the magnetic ground state to compute the excitation spectrum, with the results presented in Fig.~\ref{Fig:261}(a). Due to the extended periodicity of the spin helix, the effective lattice constant becomes $b'=6b$, and the Brillouin zone is folded by a factor of six. As a result, the number of available 
$\bfq$-points for sampling is reduced. To improve visualization, we combine data obtained from calculations on $4\times 5\times 6$ and $4\times 6\times 6$ $\bfk$-mesh in the fold-in Brillouin zone (BZ) to present in the figure. 
We first note that the magnon bandwidth is reduced to approximately 55 meV. This reduction arises because the helix spin pattern modifies the electron hopping amplitudes, which in turn affect the strength of spin exchange interactions. Overall, we identify two distinct features in the spectrum: a broad dispersive magnon band ranging from 0 to 55 meV, and a less dispersive low-energy branch that terminates near 15 meV.
\\
\indent
To further analyze the energy dependence, we integrate the spectrum over momentum as, $\bar{S}(\omega)=\sum_{q_y}S(q_y,\omega)$, which serves as an effective magnon density of states, as shown in Fig.~\ref{Fig:261}(b). A prominent low-lying feature centered around $\omega=10 \pm 5 $ meV reveals a weakly dispersive mode separated from the broader magnon spectrum. This feature may also originate from a gapped optical mode centered at $\omega=10$ meV, as predicted by the spin model in Ref.~\cite{masuda2005spin}. 
To enable comparison with experimental results, we focus on the low-energy region $\omega=0-10$ meV and identify distinct peaks, as shown in the  Fig.~\ref{Fig:261}(c). The lowest peak is fitted using a downward-opening quadratic function, and we plot both the lowest branch and the second folded-in mode. The fitting line (blue curve with light blue shaded error region) shows excellent agreement with the experimental data (red curve with pink shaded region), with a deviation of less than 1.5 meV \cite{masuda2005spin}. This result highlights the predictive power and numerical accuracy of our implementation. It is worth noting that we still observe a negative-energy state at finite momentum $q_y\approx 0.1 ~(2\pi/b')=1/60 ~(2\pi/b)$, corresponding to an extremely long unstable pitch of 60. It can be attributed to artifact of numerical resolution and safely disregarded.
\\
\indent
In conclusion, our first-principles calculations successfully reproduce the low-energy magnon dispersion in the helical spin state with high accuracy. For the broad dispersive band, the predicted bandwidth is approximately twice as large as that reported in the spin model of Ref.~\cite{masuda2005spin}. However, due to the current lack of corresponding experimental data, further validation will be required in future studies.

\section{conclusions \label{sect:conclusions}}
\vspace{-10pt} 
In this work, we present a systematic first-principles approach to compute magnetic excitations in large-scale noncollinear spin structures. The core of our formalism combines DFT with MBPT, where magnetic excitations are described as four-point correlation functions solved via the BSE. By extending the Green’s function to include off-diagonal components, we incorporate intrinsic spin transitions characteristic of noncollinear magnetic backgrounds—thereby going beyond the limitations of existing collinear frameworks.
\\
\indent
While direct implementation of the formalism faces overwhelming computational costs—primarily due to space–time integration and the need for self-consistent DFT calculations of noncollinear ground states—we introduce two key techniques to overcome these challenges. First, we utilize a Wannier function basis to reduce the dimensionality of the BSE. As an existing technique for collinear formalism, Wannier functions further provide a natural basis for defining the local spin coordinates in noncollinear system. Second, we adopt the ansatz potential method to efficiently generate noncollinear magnetic exchange potentials, thereby bypassing the extreme computational demands associated with large-scale spin textures. This combined framework enables first-principles calculations of spin-wave excitations in complex noncollinear magnetic systems on modern computational platforms.
\\
\indent
After establishing the methodology, we apply our formalism to compute the magnetic excitation spectrum in the helimagnet $\rm LiCu_2O_2$. We begin by analyzing magnetic instability in the long-wavelength limit ($\bfq=\Gamma$), where we identify three magnon modes lying below the acoustic Goldstone mode. By computing the precession pattern of each mode, we observe consistent changes in the excitation spectrum as the spin orientation in the ground state is rotated. Since the results suggest an instability that extends across multiple unit cells, we further extend the discussion to examine the magnon spectrum at finite momenta. 
In particular, we explore the relationship between exchange splitting and the momentum $q_{y,{\rm min}}$ of the lowest-energy magnon mode. This analysis reveals the nontrivial role of finite spin moment on $\rm O^{2-}$ ions in mediating effective ferromagnetic interaction for the primary spin on Cu$^{2+}$ ions. Finally, we identify the required exchange splitting that reproduces the commensurate $2\times 6\times 1$ magnetic superlattice structure, compute the corresponding magnetic excitation spectrum in the helimagnetic state, and extract the magnon dispersion. The resulting spectrum shows excellent agreement with experimental measurements, with deviations within 1.5 meV, thus validating our first-principles approach. 
\\
\indent
Overall, our work establishes a framework for predicting the magnetic excitation in a large-scale noncollinear magnet. Beyond the effective spin model approach, this formalism treats all electrons on an equal footing such that it is suitable to investigate materials containing multi-magnetic ions and is applicable to either $p$, $d$, or $f$ electron system, while the effect of itinerant electron is also properly included. For future development, we aim to generalize the formalism to the strongly correlated regime and investigate the nonperturbative effect in magnetic response.

\begin{acknowledgments}
\vspace{-10pt}
We acknowledge the financial support by Grant-in-Aids for Scientific Research (JSPS KAKENHI) Grant Numbers JP25H01252, JP23H04869, JP23H04519, and JP25H01506, 
JST-CREST No.~JPMJCR23O4, JST-ASPIRE No.~JPMJAP2317,  RIKEN TRIP initiative (RIKEN Quantum, Advanced General Intelligence for Science Program, Many-body Electron Systems), and MEXT as ``Program for Promoting Researches on the Supercomputer Fugaku'' (Project ID: JPMXP1020230411).
\end{acknowledgments}

\appendix

\section{Computational details \label{Append:detail}}
\vspace{-10pt}
We carry out the DFT calculation using the {\sc quantum espresso} package \cite{giannozzi2009quantum} with norm-conserving scalar-relativistic pseudopotential generated by PSEUDO-DOJO package \cite{van2018pseudodojo}, with the PBEsol functional under the GGA approximation \cite{perdew1996generalized,perdew2008restoring}. We adopt the  lattice constant, $A= 5.59$ \AA, $B = 2.82$ \AA, and $C=12.14$ \AA~for the orthorhombic structure, optimized in FM state, which is slightly smaller shorter than the experiment value, $A_{\rm exp.}= 5.73$ \AA, $B_{\rm exp.} = 2.86$ \AA, and $C_{\rm exp.}=12.42$ \AA , possibly due to the different magnetic structure. 
For the electronic ground state and exchange potential, we conduct the noncollinear self-consistent field (scf) calculation without spin-orbit coupling on a $4\times 8\times 4$ $\bfk$-grid with an energy cut-off 95 Ry, which corresponds to a $72\times 36\times 144$ fast Fourier transform grid. We further take the scalar part of the exchange potential and make non-scf calculation on $8\times 30\times 6$ and $8\times 36\times 6$ $\bfk$-grid to obtain the Kohn-Sham orbitals, from which we use the Wannier90 package \cite{mostofi2014updated,pizzi2020wannier90} to construct the localized Wannier orbitals with initial guess of $d$ orbitals on Cu atom and $p$ orbitals on O atom.
\\
\indent
For the static Coulomb interaction, we use the RESPACK package \cite{nakamura2021respack} to compute the screened dielectric constant with random phase approximation (RPA) with 180 spin polarized band and 5 Ry plane wave cut-off on a $10\times 20\times 5$ $\bfk$-grid, while for consistency we manually apply a 430 meV rigid shift to the band gap to match with the required value as shown in Fig.~\ref{Fig:EXvsDISP}(b). We later interface the Wannier90 and RESPACK to compute the on-site Coulomb interaction among Wannier orbitals.
\\
\indent
To estimate the band gap, we use the Yambo code \cite{sangalli2019many} to carry out the GW calculation \cite{hybertsen1986electron}. The dynamical Coulomb screening is obtained under the plasmon-pole approximation with 180 bands and a plane wave cut-off 5 Ry is applied, while 104 valence band and 116 conduction band used in GW energy calculation, for which we didn't seek a convergent value. In order to correctly capture the insulating nature when computing the screening, instead of the generally adopted one-shot $G_0W_0$ calculation, we report the result obtained from one more iteration, i.e. $ G_1W_1$.
\\
\indent
When computing the magnetic response, as discussed in the Section \ref{sect:application}, we restrict the scattering channel by using the $d_{xy}$ orbitals on the Cu$^{2+}$ ions and $p_{x,y}$ orbitals on the O$^{2-}$ orbitals with totally 20 Wannier orbitals in a unit cell. When computing the Green's function of Eq.~(\ref{Eq:Kab_general}), since the Wannier coefficient $U_{Nm\bfk}$ is smooth function in respective to $\bfk$, we adopt the double grid method to improve the convergence. Explicitly, we 
focus on the component with strong momentum dependence and make following replacement:
\begin{eqnarray}
    &&\frac{f_{m\bfk+\bfq}-f_{n\bfk}}{\w-(\e_{m\bfk+\bfq}-\e_{n\bfk})+i\eta}\nn\\
    &&~~~\rightarrow
    \frac{1}{N_{\d \bfk}}\sum_{\d \bfk}
    \frac{f_{m\bfk+\d \bfk+\bfq}-f_{n\bfk+\d \bfk}}{\w-(\e_{m\bfk+\d \bfk+\bfq}-\e_{n\bfk+\d \bfk})+i\eta}
\end{eqnarray}
where the $\d \bfk$ takes account for the $\bfk$-point on the fine grid around the target $\bfk$-point on the course grid, and the eigenenergies are obtained efficiently by the Wannier interpolation method. In this work, we effective extend the $4\times 5\times 6$ grid to $12\times 15\times 6$ and $4\times 6\times 6$ grid to $12\times 18\times 6$ in the fold-in BZ with a converged accuracy for the magnon energy up to 1 meV.

\section{Details on constructing the ansatz potential \label{Append:AP-method}}
\vspace{-10pt}
\begin{figure}[t]
    \centering
    \includegraphics[width=1.0\linewidth]{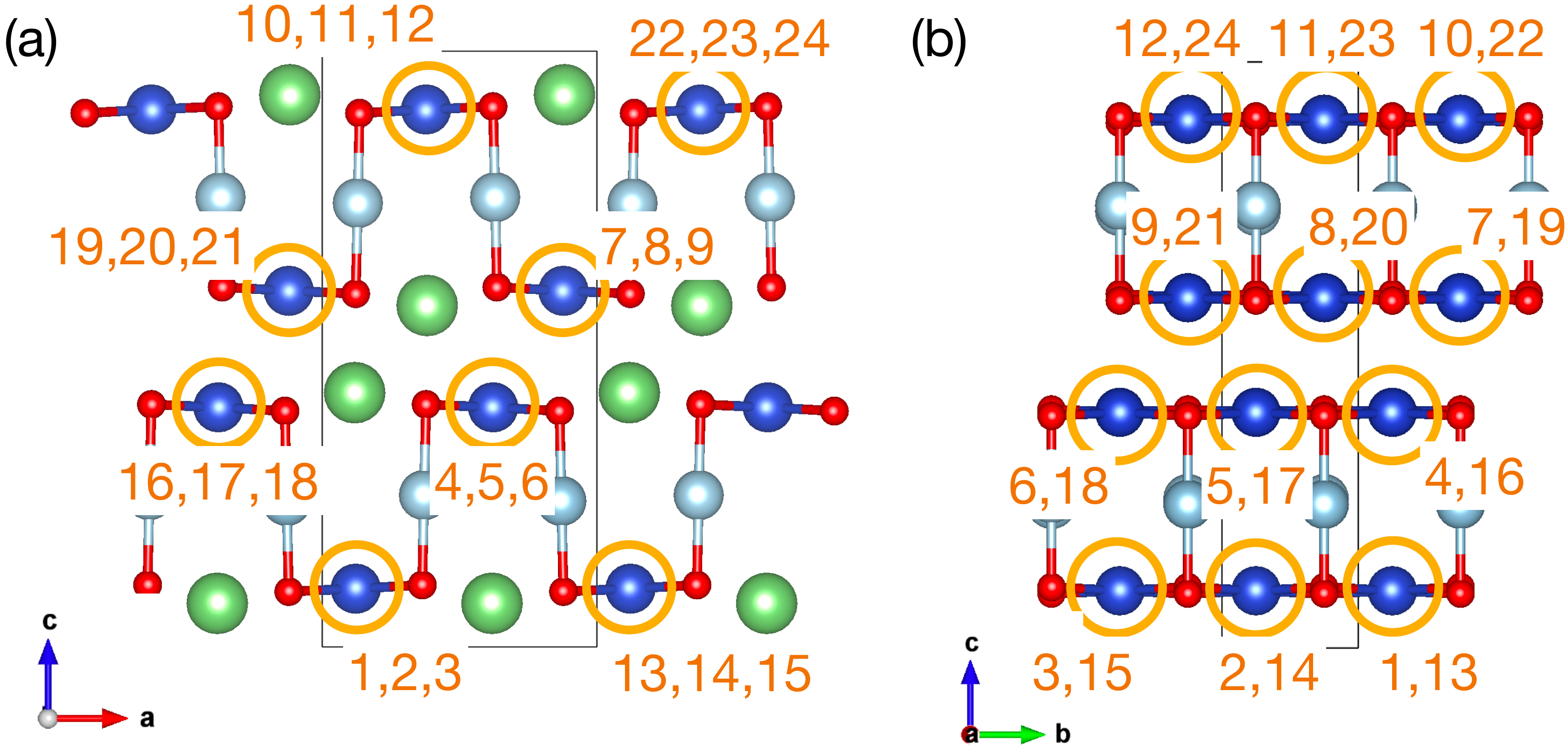}
    \caption{24 spins on the encircled magnetic atoms are assumed to determine the magnetic part of exchange potential in the unit cell. (a) Side view of the crystal along $b$-direction. (b) Side view of the crystal along $a$-direction.  }
    \label{Fig:AP_LCO}
\end{figure}
In this section, we present the detail to apply the ansatz potential method \cite{chen2025topological} for $\rm LiCu_2O_2$. We set the ansatz function for the magnetic part of exchange potential as the Heisenberg type with noncollinear component:
\begin{eqnarray}
    \bfB_{S}(\bfx)=\sum_a  \overleftrightarrow{\bf J}_a(\bfx)\cdot {\bf S}^a.
    \label{Eq:anastz}
\end{eqnarray}
Since we want to focus on the in-plane helix structure, we need only the $x$, $y$ component for the $\bfB_s$ and $\bfS$. Therefore, $\overleftrightarrow{\bf J}_a(\bfx)$ becomes a tensor function of $2\times 2$, spanning over the unit cell. The $a$ denotes the magnetic atom index. In this study, we include totally 24 spins in Eq.~(\ref{Eq:anastz}), whose positions are presented in Fig.~\ref{Fig:AP_LCO}. In order to obtain the fitting data, we carried out 12 independent DFT calculation on $1\times 2\times 1$ supercell, with arbitrary assigned in-plane spin orientation as initial guess, to acquire the self-consistent potentials, which, along with symmetry operation, provide totally 192 data points. The resultant $\overleftrightarrow{\bf J}_a(\bfx)$ is further refined by the crystal symmetry. It is worthy to note that compared to the $f$-electron system, in $d$-electron system the energy of arbitrary orientated spin structure is in general not a local minimum. Therefore, to stabilize the calculation, we need to introduce tiny energy penalty to confine the spin close to the initial guess. The result can reproduce the DFT band structure with an error less than 20 meV.

\section{Extending primitive cell Wannier basis to supercell \label{sect:unit2super}}
In this section, we provide a supplementary discussion on extending the Wannier basis and Hamiltonian from a primitive cell to a supercell representation in order to incorporate the exchange potential induced by the spin texture as in Eq.~(\ref{Eq:Hsp}). Although Wannierization can, in principle, be performed directly within the supercell, this approach is computationally expensive due to the increased number of Kohn–Sham states and charge densities that must be calculated over a larger configuration space. Additionally, much of this information is redundant and can be reduced by leveraging translational symmetry. Moreover, since the Wannier functions constructed in the primitive cell and those obtained from a supercell are expected to be equivalent up to integer translations among cells, extending the Wannier basis from the primitive cell offers a more efficient and physically consistent alternative to direct supercell construction.
\\
\indent
In the standard approach, Wannier orbitals for a primitive cell system are constructed as linear combination of Bloch wave functions across bands and over the BZ, while minimizing the total quadratic spreads of each orbital around  its center \cite{mostofi2014updated,pizzi2020wannier90}. These Wannier functions are related to the Bloch states through:
\begin{eqnarray}
    &&|w_N \bfR_{\rm p}\>=\frac{1}{N_{\bfk}}\sum_{\bfk} e^{i\bfk\cdot(\hat{x}-\bfR_{\rm p})}|u^{\rm (W)}_{N\bfk}\>
    \nn\\
    &&~~~~~~~~~~~~=\frac{1}{N_{\bfk}}
    \sum_{n\bfk} e^{i\bfk\cdot(\hat{x}-\bfR_{\rm p})}U^*_{Nn\bfk}|u_{n\bfk}\>
    .\nn\\
    &&|u_{n\bfk}\>=\sum_{N}U_{Nn\bfk}|u^{\rm (W)}_{N\bfk}\>\nn\\
    &&~~~~~~~~~~~~=\sum_{N\bfR_{\rm p}} e^{-i\bfk\cdot(\hat{x}-\bfR_{\rm p})}U^*_{Nn\bfk}|w_N \bfR_{\rm p}\>
    \label{Eq:uwk}
\end{eqnarray}
where $|w_N \bfR_{\rm p}\>$ denotes the $N$-th Wannier orbital centering in the primitive cell locating at $\bfR_{\rm p}$, and $|u^{\rm (W)}_{N\bfk}\>$ is called the Bloch function in Wannier Gauge. Once the Wannier functions are constructed, the momentum-space Hamiltonian $H^{\text{non-mag}}(\bfk)$ can be transformed into a real-space tight-binding representation via the matrix elements $H^{\text{non-mag}}_{NM}(\bfR_{\rm p})=\<w_{N} {\bf 0}_{p}|H^{\text{non-mag}}|w_{M} {\bfR_{\rm p}}\>$.
Owing to the exponential localization of Wannier orbitals, the hopping terms decay rapidly with distance, allowing the Hamiltonian to be accurately reconstructed by including only matrix elements within a spatial cutoff $|\bfR_{\rm p}|<{\rm R_{cut}}$. The real-space and momentum-space Hamiltonians are connected through a Fourier transform:
\begin{eqnarray}
    H^{{\text{non-mag}}(W)}(\bfk)=\sum_{\bfR_{\rm p}}e^{i\bfk\cdot\bfR_{\rm p}}H^{\text{non-mag}}(\bfR_{\rm p})
    \label{Eq:Hwk_prime}
\end{eqnarray}
where the superscript $\rm (W)$ indicates that $H^{\rm (W)}(\bfk)$ is expressed in the Wannier gauge.
\\
\indent
\begin{figure}[t]
    \centering   \includegraphics[scale=0.2]{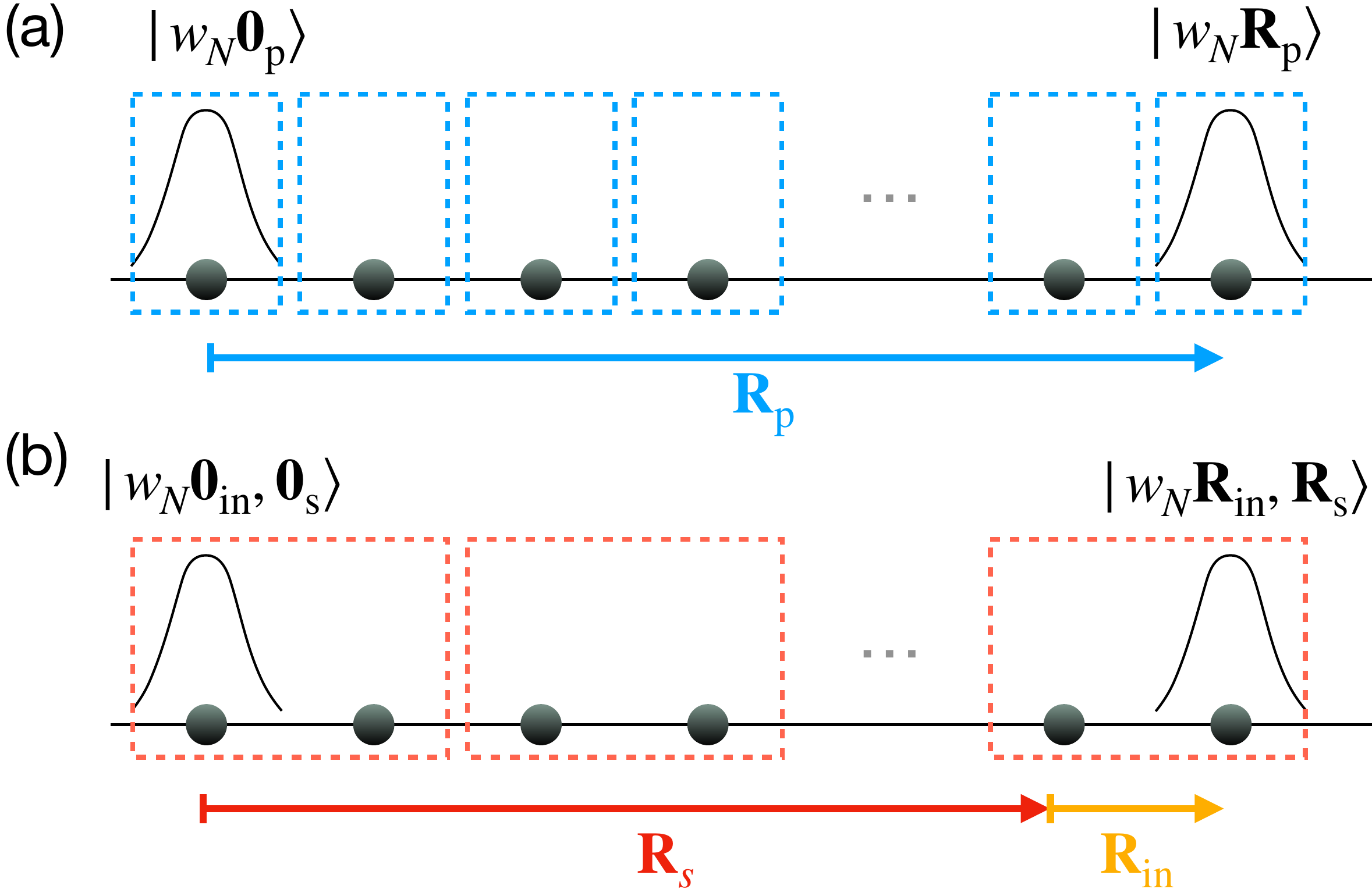}
    \caption{(a) Wannier orbitals in the primitive cell representation (b) Wannier orbitals in the supercell representation. }
    \label{Fig:RRR}
\end{figure}
Starting from the standard Wannier function construction, we extend the Hamiltonian from the primitive cell to a supercell of size $S_1\times 1\times 1$. In this supercell, the number of independent Wannier orbitals increases by $S_1$ times. To label these orbitals, we introduce the following changes on notation:
\begin{eqnarray}
    |w_N \bfR_{\rm p}\> \rightarrow |w_N \bfR_{\rm in}, \bfR_{\rm s}\> 
\end{eqnarray}
where $\bfR_{\rm s}$ denotes the cell position vector of the supercell, $\bfR_{\rm in}$ is an internal translation within the $S_1\times 1\times 1$ supercell such that they are related to the original primitive lattice vector by $\bfR_{\rm p}=\bfR_{\rm s}+\bfR_{\rm in}$. The pair $(w_N, \bfR_{\rm in})$ thus serves as a composite index for labeling Wannier orbitals within the supercell unit cell (see Fig.~\ref{Fig:RRR}). With this labeling, the tight-binding Hamiltonian in the supercell can be constructed using the same prescription:
\begin{eqnarray}
    &&H_{NM}(\bfR_{\rm p})=\<w_N{\bf  0}_p|H|w_M\bfR_{\rm p}\>\nn\\
    &&\rightarrow H^{\rm s}_{N\bfR_{\rm in,1},M\bfR_{\rm in,2}}(\bfR_{\rm s})=\<w_N \bfR_{\rm in,1},{\bf 0}_{\rm s}|H|w_M\bfR_{\rm in,2},\bfR_{\rm s}\>\nn\\
    \label{Eq:HR_super}
\end{eqnarray}
which follows the translation symmetry:
\begin{eqnarray}
   &&\<w_N \bfR_{\rm in,1},{\bf 0}_{\rm s}|H|w_M\bfR_{\rm in,2},\bfR_{\rm s}\>
   \nn\\
   &&=~
   \<w_N {\bf 0}_{\rm s},{\bf 0}_{\rm s}|H|w_M\bfR_{\rm in,2}-\bfR_{\rm in,1},\bfR_{\rm s}\>\nn\\
   &&=~~~~~\<w_N{\bf 0}_{\rm p}|H|w_M\bfR_{\rm s}+\bfR_{\rm in,2}-\bfR_{\rm in,1}\>.
\end{eqnarray}
In Eq.~(\ref{Eq:HR_super}), we use the superscript ``$\rm s$" in $H^{\rm s}$ to emphasize that the Hamiltonian is constructed in a super cell system. 
When we choose the supercell to match the periodicity of the spin texture, Eq.~(\ref{Eq:HR_super}) becomes compatible with Eq.~(\ref{Eq:Hsp}) to incorporate the magnetic interaction from the spins, 
\begin{eqnarray}
    &&H^{\rm s, mag}_{N\bfR_{\rm in,1}M\bfR_{\rm in,2}}(\bfR_{\rm s})=H^{\rm s, \text{non-mag}}_{N\bfR_{\rm in,1}M\bfR_{\rm in,2}}(\bfR_{\rm s})+\nn\\
    &&~~\<w_N\bfR_{\rm in,1},{\bf 0}_{\rm s}|\bfB_{\rm KS}(\bfx)|w_M\bfR_{\rm in,2},\bfR_{\rm s} \>
\end{eqnarray}
Finally, the momentum-space Hamiltonian can be obtained via Fourier transform, for $\bfk_{\rm s}$ being the wave vector in the folded-in Brillouin zone of the supercell:
\begin{eqnarray}
    H^{\rm s, mag~(W)}(\bfk_{\rm s})=\sum_{\bfR_{\rm s}}e^{i\bfk_{\rm s}\cdot\bfR_{\rm s}}H^{\rm s,mag}(\bfR_{\rm s}).
    \label{Eq:Hk_commense}
\end{eqnarray}
After sovling the Hamiltonian, the corresponding eigenvalues and eigenvectors can serve as the band energies and mixing matrices required for constructing the two-particle Green’s function in Eq.~(\ref{Eq:Kab_general}).
\\
\indent
In addition, when computing the magnetic response, Eq.~(\ref{Eq:qW}) requires the Bloch wave functions expressed in the Wannier gauge. These wave functions can be reconstructed as:
\begin{eqnarray}
    &&|u^{\rm s(W)}_{(N\bfR_{\rm in})\bfk_{\rm s}}\>=\sum_{\bfR_{\rm s}}e^{-i\bfk_{\rm s}\cdot(\hat{x}-\bfR_{\rm s})}|w_N\bfR_{\rm in},\bfR_{\rm s}\>\nn\\
    &&
=\sum_{\bfR_{\rm s}}e^{-i\bfk_{\rm s}\cdot(\hat{x}-\bfR_{\rm s})}
    \frac{1}{N_{\bfk_{\rm p}}}
    \sum_{\bfk_{\rm p}} e^{i\bfk_p\cdot(\hat{x}-\bfR_{\rm s}-\bfR_{\rm in})}|u^{\rm (W)}_{N\bfk_{\rm p}}\>\nn\\
    &&=\frac{1}{N_{\bfk_{\rm s}}S_1}
    \sum_{\bfR_{\rm s}\bfk'_{\rm s}}\sum_{s_1=0}^{S_1-1}
    e^{i(\bfk_{\rm s}'-\bfk_{\rm s}+s_1\bfb)\cdot(\hat{x}-\bfR_{\rm s})}\nn\\
    &&~~~~~~~~~~~~~~~~~~~~~~~~~~~\times 
    e^{-i(\bfk'_{\rm s}+s_1\bfb)\cdot\bfR_{\rm in}}
    |u^{\rm (W)}_{N\bfk_{\rm s}'+s_1\bfb}\>\nn\\
    &&=
    e^{-i\bfk_{\rm s}\cdot\bfR_{\rm in}}
    \frac{1}{S_1}
    \sum_{s_1=0}^{S_1-1}
    e^{is_1\bfb\cdot(\hat{x}-\bfR_{\rm in})}
    |u^{\rm (W)}_{N\bfk_{\rm s}+s_1\bfb}\>.
\end{eqnarray}
Here we introduce the relation $\bfk_{\rm p}=\bfk_{\rm s}+s_1\bfb$, where $\bfk_{\rm p}$ is the wave vector in the unfolded (primitive-cell) BZ, and $\bfb$ is a reciprocal-space vector that accounts for the zone folding,  satisfying ${\rm Mod}_{2\pi}(\bfb\cdot\bfR_{\rm s})=0$.

\bibliography{Mag_ref}
\end{document}